\documentclass{appolb}
\usepackage{epsfig}
\def\GeV{{\rm GeV}}
\def\MeV{{\rm MeV}}

\begin{document}
\pagestyle{plain}
\title{The Structure Function Working Group Summary%
\thanks{Presented at DIS2002, Cracow, 30 April - 4 May 2002}%
}
\author{Vladimir Chekelian (Shekelyan)
\address{MPIM (Munich) and ITEP (Moscow)}
\and
Amanda Cooper-Sarkar
\address{Oxford University}
\and
Robert Thorne
\address{Cambridge University}
\thanks{Royal Society University Research Fellow}
}
\maketitle
\begin{abstract}
A summary of the experimental and theoretical presentations
in the Structure Function Working Group
on the proton and photon unpolarized structure functions is given.
\end{abstract}

\section{Introduction}
In this report we summarize new results on 
lepton-nucleon deep inelastic scattering (DIS) from 
the fixed target experiments NuTeV and Hermes,
and from the $ep$ collider HERA in the range of
four-momentum transfer squared, $Q^2$, 
between 0.35 and 30000~GeV$^2$.
Recent next-to-leading-order (NLO) QCD fits and uncertainties 
of the parton distribution functions (PDFs) are discussed.
In the theory part of this summary, developments in lattice QCD, 
saturation type effects and colour glass condensates, 
$k_T$-factorization, and collinear factorization are surveyed.

\section{DIS results on the proton structure functions}
\subsection{The first NuTeV results on $F_2$}
\begin{figure}[ht] 
\begin{center}
\epsfig{file=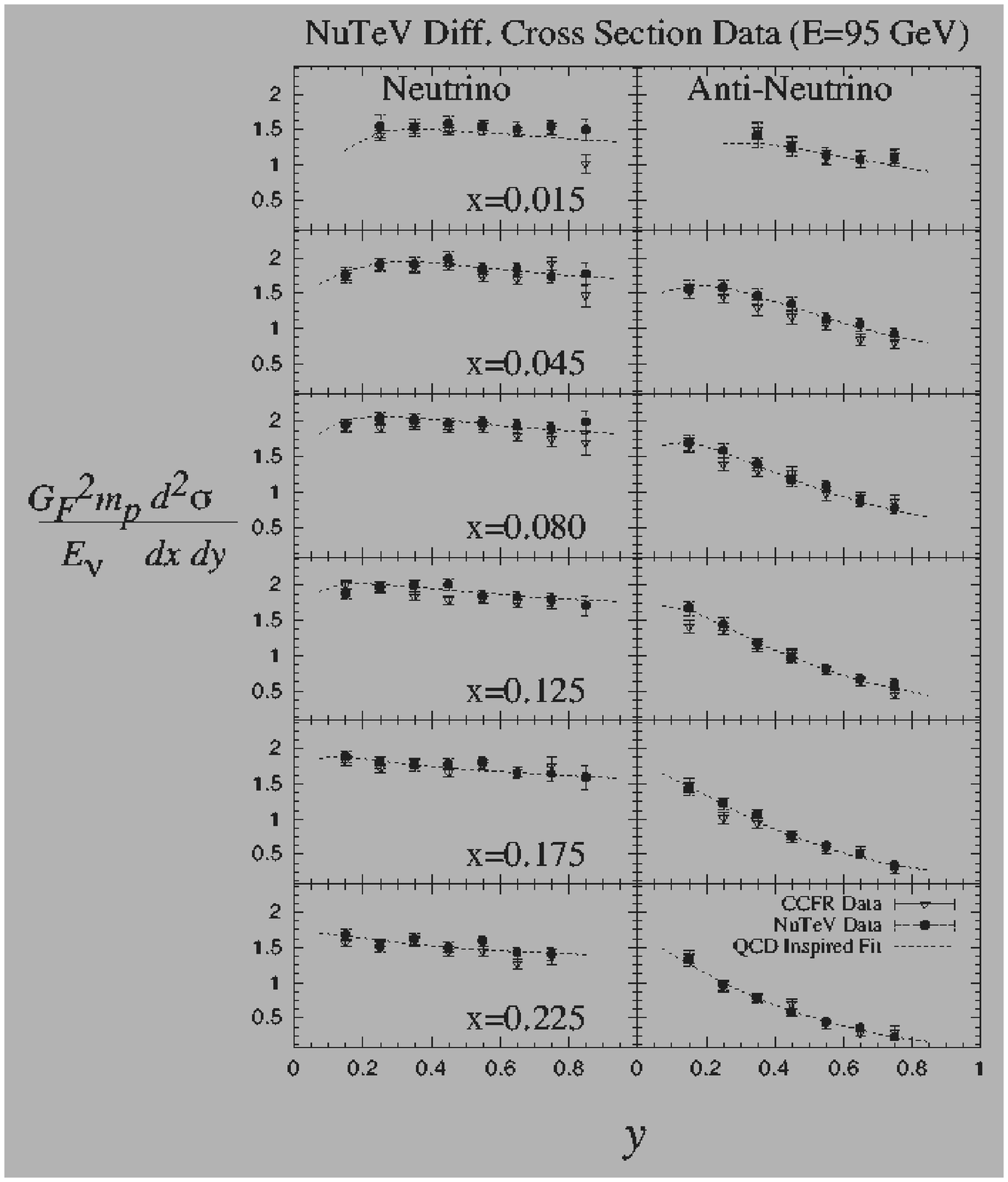,width=6.5cm} 
\epsfig{file=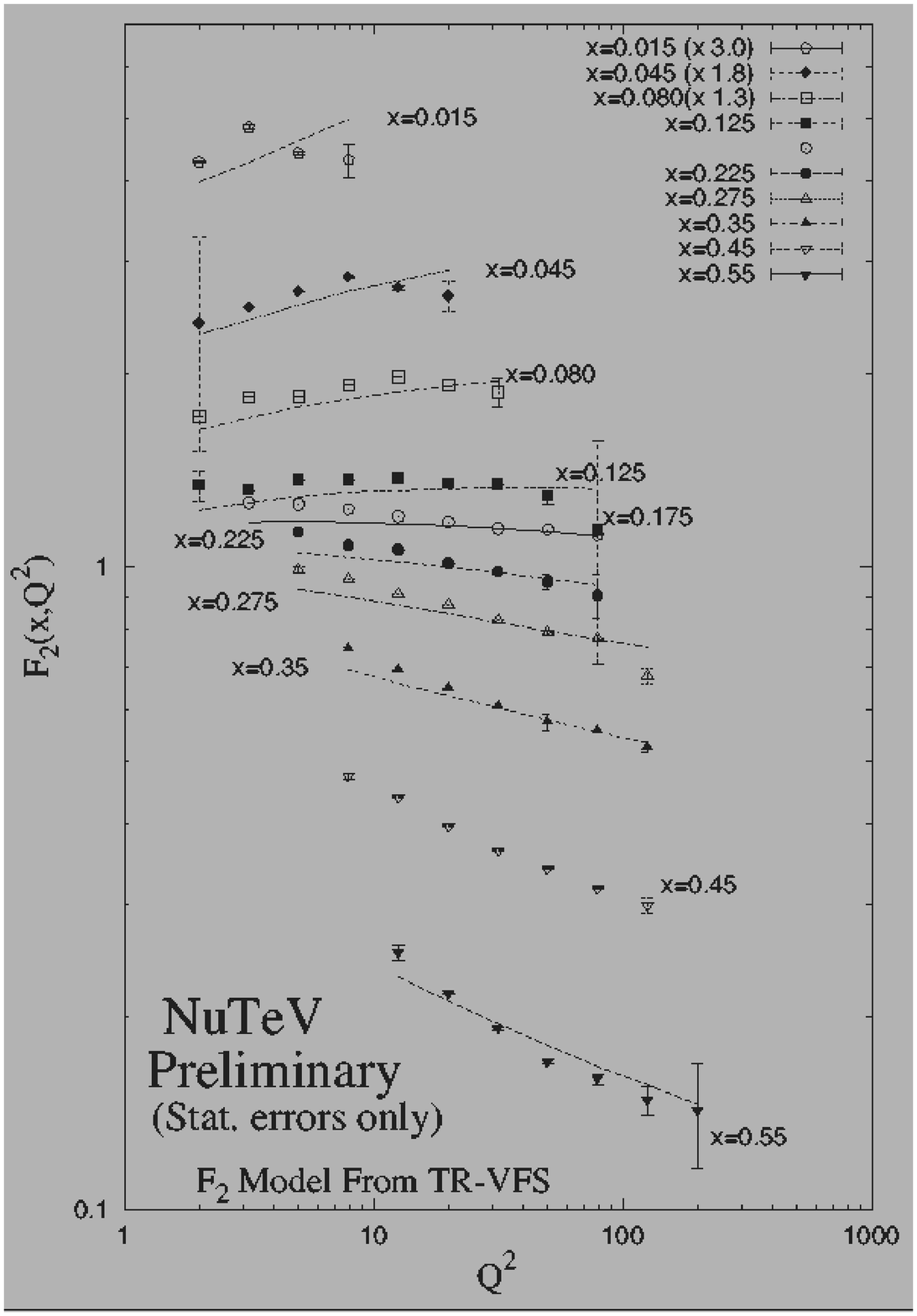,width=5.5cm} 
\caption     { \sl 
$\nu(\overline{\nu})$-nucleon cross sections (left) and 
$F_2(x,Q^2)$ (right) from NuTeV. }
\label {nutev}
\end{center}
\end{figure}

The $\nu(\overline{\nu})$-nucleon cross sections
from the NuTeV neutrino experiment
were presented by R. Bernstein~\cite{nutev}.
In Fig.\ref{nutev} (left)
the results are compared with the former CCFR data~\cite{ccfr} 
as function of the inelasticity $y$.
The measurements are in a good agreement apart from
$x=0.45$ where the $\overline{\nu}$ results of CCFR are
systematically lower.
In contrast to CCFR the NuTeV experiment uses 
very clean $\nu$ and $\overline{\nu}$ beams provided by
a Sign Selected Quadrupole Train where 
the charge of the parent $\pi, K$ of the neutrinos
can be selected.
The admixtures of the wrong neutrino type is 
$3 \cdot 10^{-4}$ for $\nu$ and $4 \cdot 10^{-3}$ for
$\overline{\nu}$ beams.
The energy scale uncertainties for muons and hadrons
are also improved compared to CCFR with
0.8$\%$ for muon (goal 0.3$\%$) and 0.4$\%$ for hadrons.
For CCFR both uncertainties were 1$\%$.

The sum of $\nu$ and $\overline{\nu}$ cross sections 
depends on $F_2$, $R$, the ratio
$\sigma_L/\sigma_T$ of longitudinal to transverse cross sections, 
and $\Delta x F_3 = x F_3^{\nu}-x F_3^{\overline{\nu}}=4x(s-c)$
which is sensitive to heavy quark densities.
All three functions cannot be derived from the data
simultaneously because of strong correlations among
corresponding parameters.
The first NuTeV results on $F_2(x,Q^2)$, 
shown in Fig.\ref{nutev} (right), were obtained using 
the world knowledge on $R$ and $\Delta x F_3$ deduced
from the $y$ dependence of the cross sections.

\subsection{New HERMES results on nuclear effects in DIS}
\begin{figure}[ht] 
\begin{center}
\epsfig{file=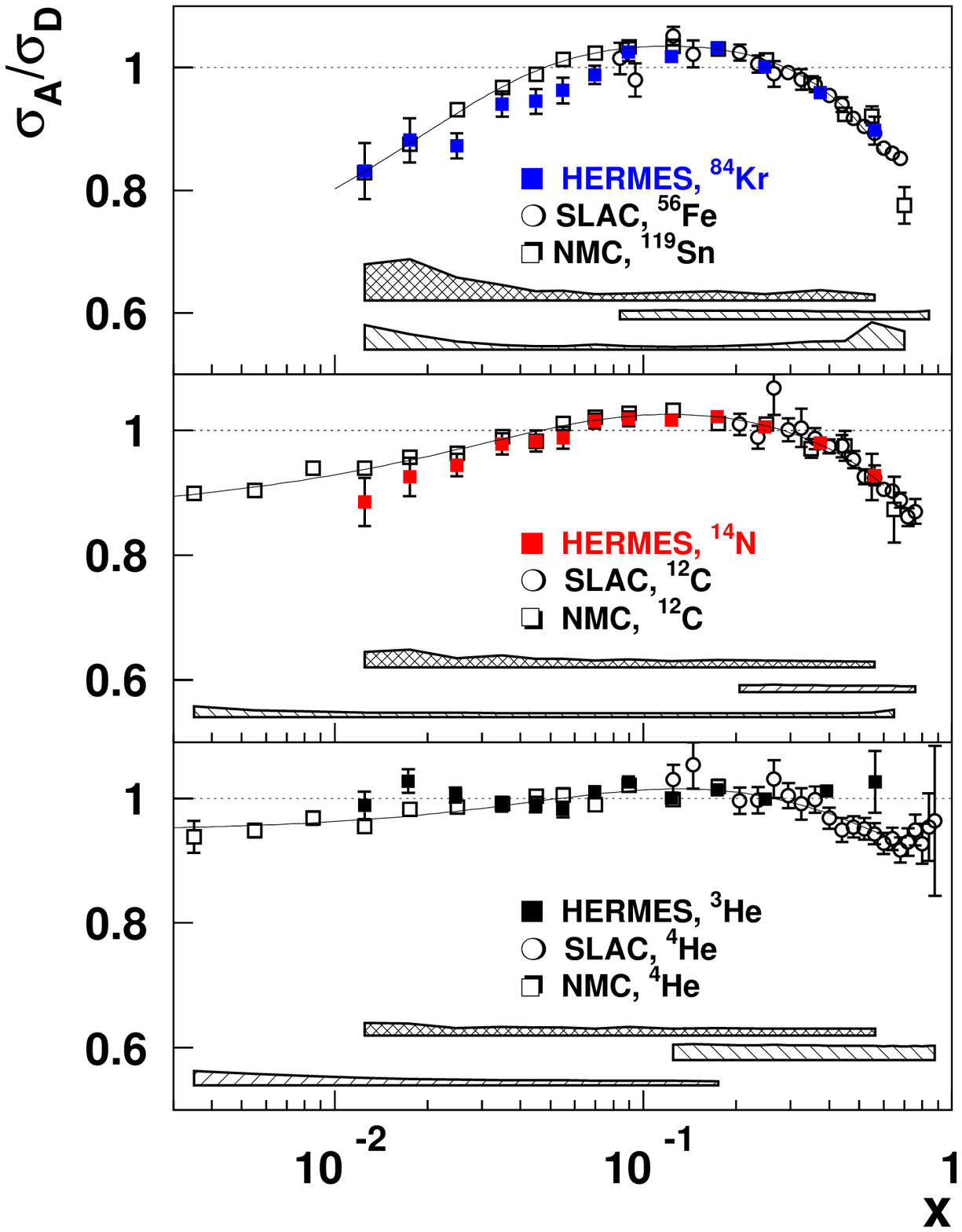,width=6.2cm} 
\epsfig{file=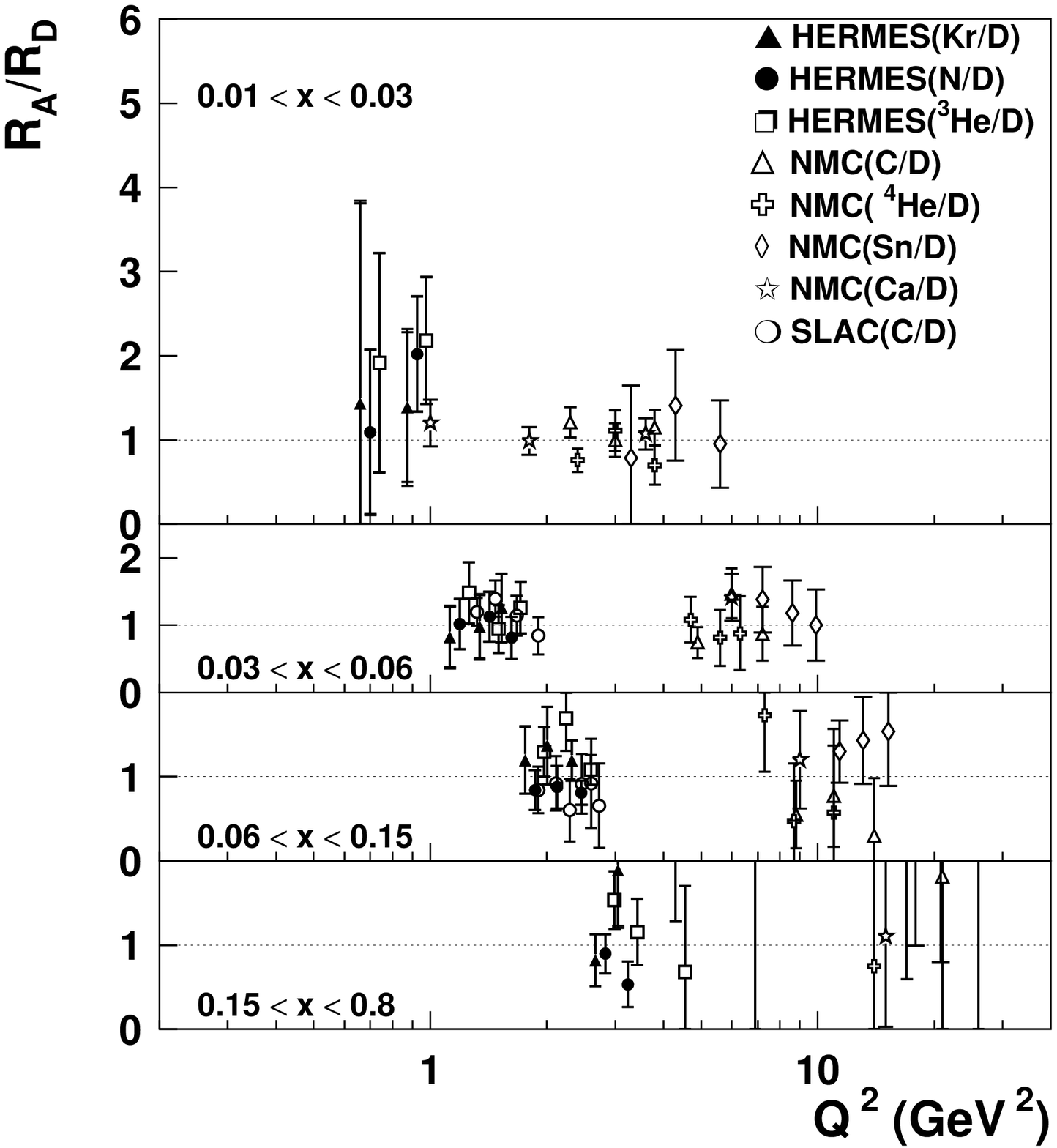 ,width=6.2cm} 
\caption { \sl 
Ratio of isoscalar DIS Born cross sections 
for several nuclei with respect to deuterium (left),  
ratio $R_A/R_D$ (right).}
\label {hermes}
\end{center}
\end{figure}

DIS cross section ratios 
for positrons of 27.5 GeV on helium-3, nitrogen and krypton 
with respect to deuterium were measured by the HERMES collaboration
(presented by A. Bruell~\cite{hermes}).
The helium-3 and nitrogen data were already published~\cite{oldhermes}.
Recently, those data were found to suffer
from an A-dependent tracking efficiency of the
HERMES spectrometer, which was not recognised in the
previous analysis. The resulting correction of the
cross section ratios is significant at low values of $x$ and $Q^2$ 
and substantially changes the interpretation of those data. 
The data corrected for this effect are shown 
in Fig.\ref{hermes}. They are in agreement with
previous measurements of NMC and SLAC.
Values for the ratio $R_A/R_D$ 
have been derived from the the $y$ dependence of the data
and are found to be consistent with unity. 

\subsection{HERA results at low and medium $Q^2$}
\label{sec:lowq2}
\begin{figure}[ht] 
\begin{center}
\epsfig{file=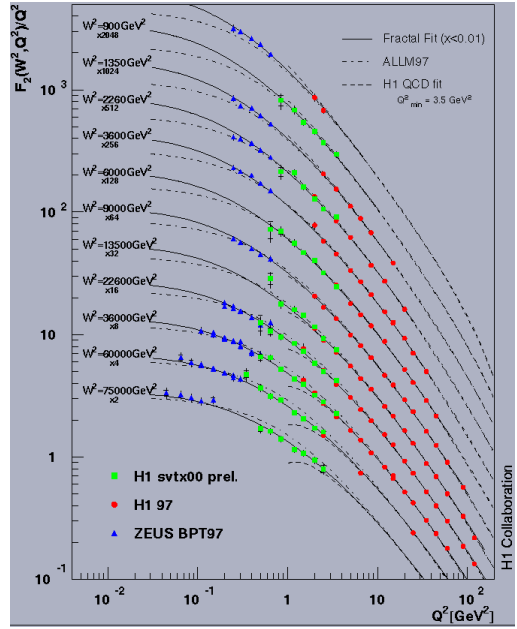 ,width=6.7cm} 
\caption { \sl 
$F_2/Q^2 \sim \sigma_{tot}(\gamma^*p)$ 
as function of $Q^2$ for different $W$,
the invariant mass of the hadronic final state.
Grey squares at $0.35 < Q^2 < 3.5$~GeV$^2$ correspond to the new H1 results. 
}
\label {xsecgp}
\end{center}
\end{figure}

T. Lastovicka~\cite{svx00} presented 
new H1 data on $F_2(x,Q^2)$ at very low $x$ and 
$0.35 < Q^2 < 3.5$~GeV$^2$
in the transition region from the non-perturbative QCD to 
the DIS domain, see Fig.\ref{xsecgp}.
The data were taken in 2000 in a special run 
with the interaction vertex shifted by 70 cm in the proton beam direction,
thereby accessing lower $Q^2$ than at the nominal vertex position.
The luminosity was increased 
by about a factor of four as compared to the initial
shifted vertex run in 1995
which lead to the first H1~\cite{zsh95h1} and ZEUS~\cite{zsh95zeus} 
data on the proton structure function  in the low $Q^2$ domain.
 
\begin{figure}[ht] 
\begin{center}
\epsfig{file=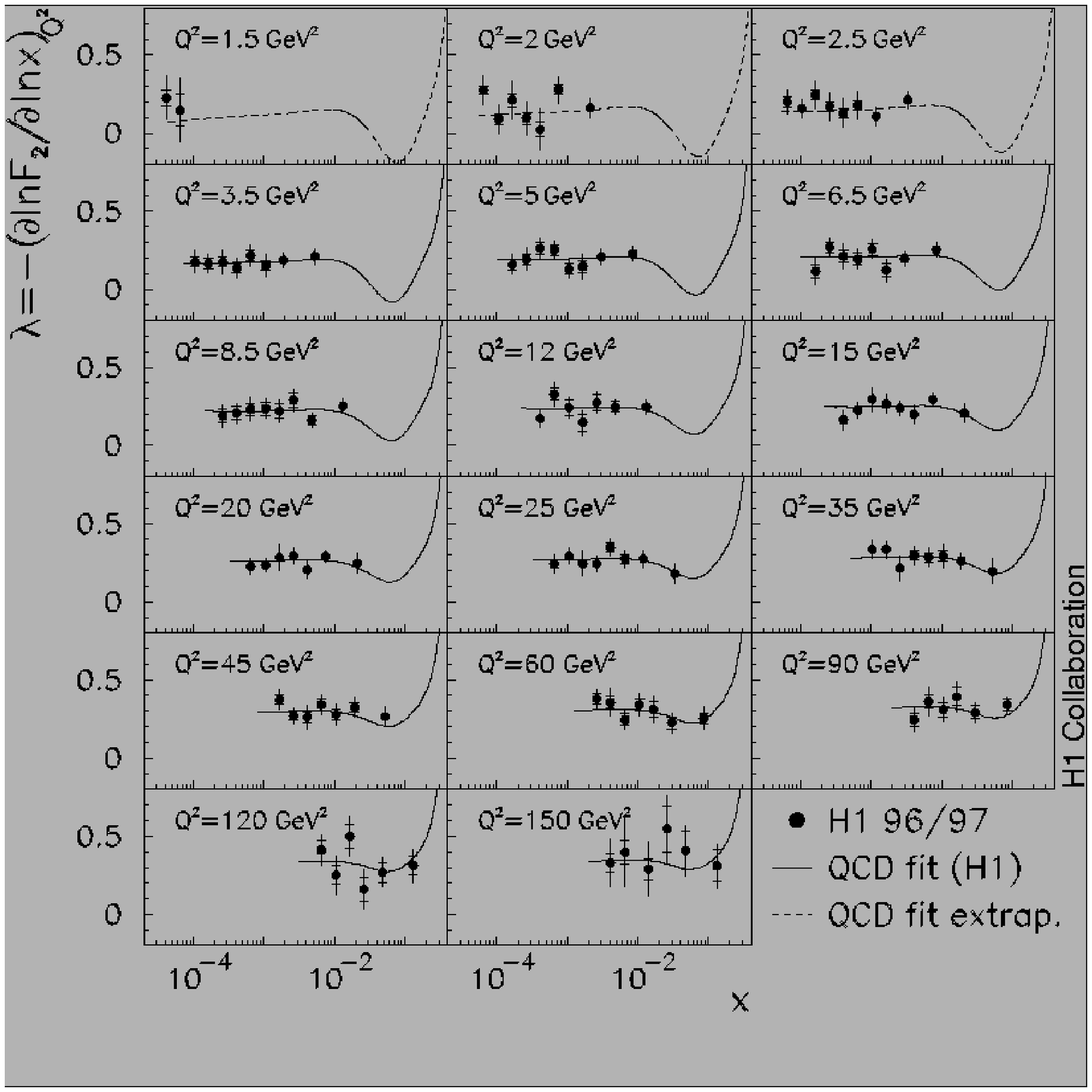,width=5.8cm} 
\epsfig{file=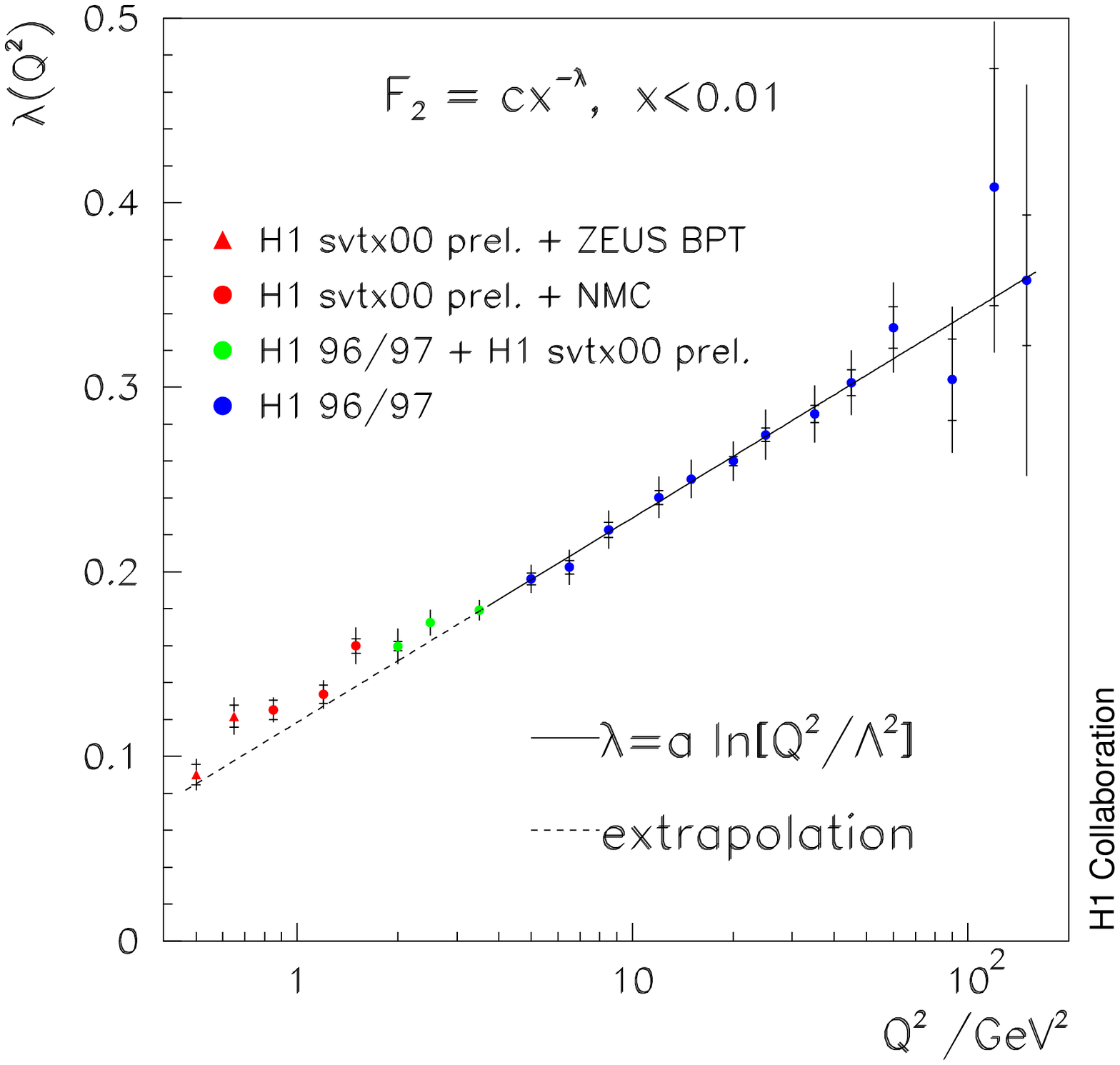,width=6.3cm} 
\epsfig{file=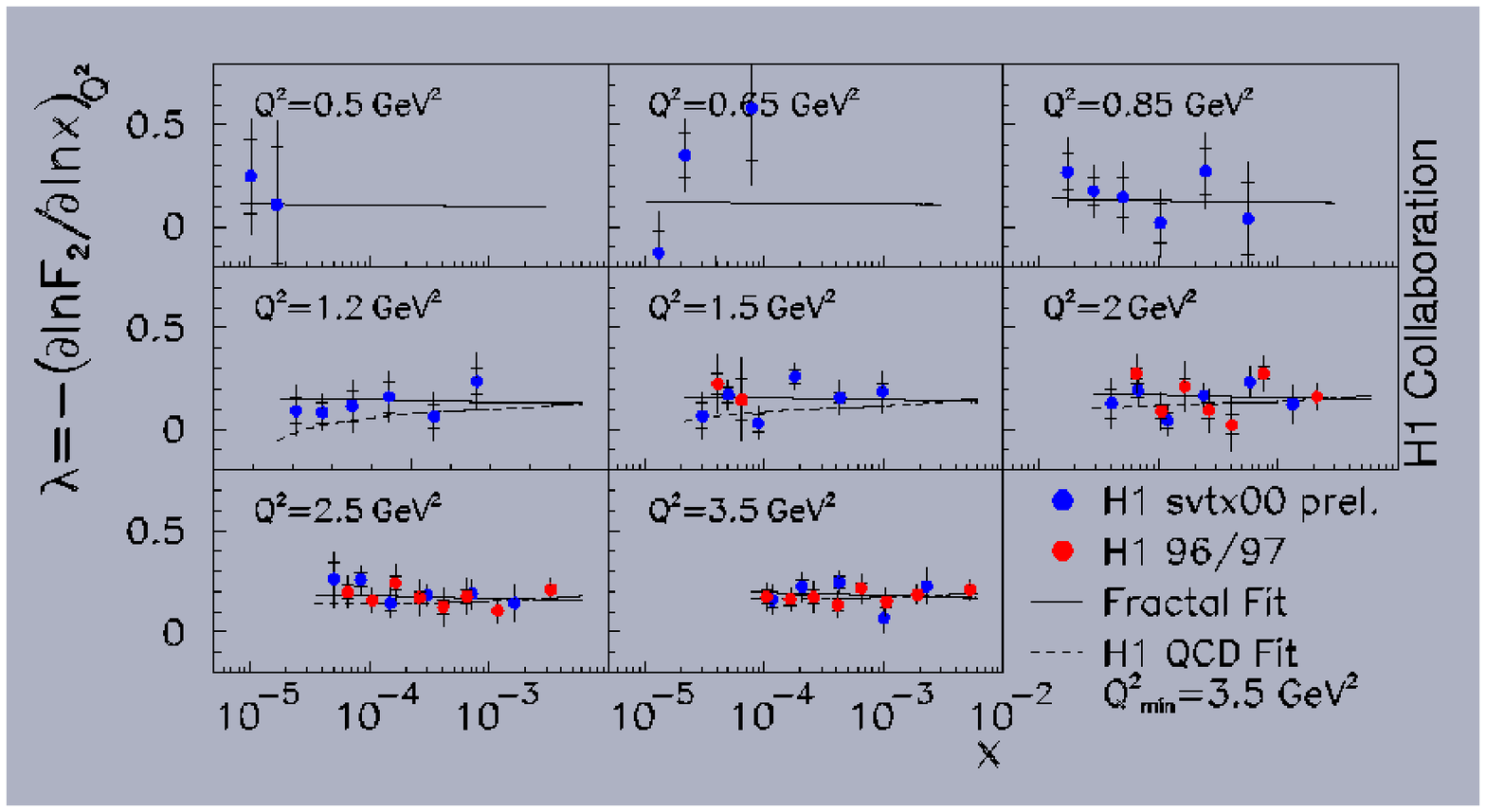,width=5.8cm} 
\epsfig{file=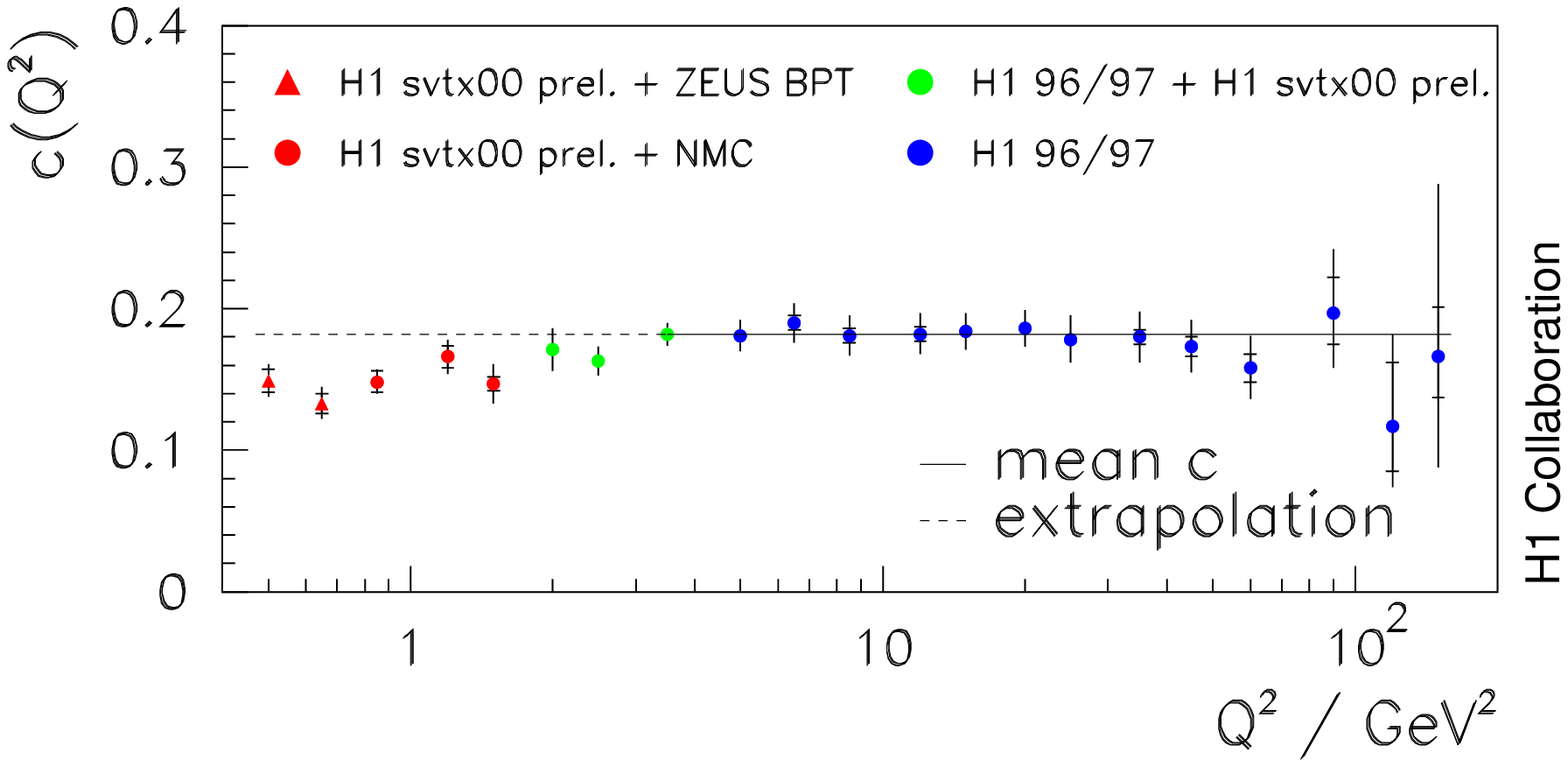,width=6.3cm} 
\caption { \sl Local derivative 
$\lambda = -(\partial \ln F_2 / \partial \ln x  )_{Q^2}$ (left)
and fitted values of $\lambda(Q^2)$ and $c(Q^2)$ (right).}

\label {lambda}
\end{center}
\end{figure}

The steep rise of the proton structure function $F_2$ towards
small $x$ was first observed in 1993 in the HERA data~\cite{hera93}.
In perturbative QCD this rise corresponds to an increase of 
the gluon density and is expected to slow down at highest energies 
(small $x$) due to gluon-gluon interactions.
Meanwhile the precision of the $F_2$ data is much improved and 
the rise is studied in great detail. 
J. Gayler~\cite{lambda} presented the local derivative 
$\lambda = -(\partial$ ln $F_2 / \partial$ ln $x  )_{Q^2}$ 
based on the new H1 $F_2$ data~\cite{svx00}
and published precision H1 data~\cite{h1lowq2}.
The $x$ and $Q^2$ dependence of $\lambda$ is shown 
in Fig. \ref{lambda} (left). The derivative is constant for
fixed $Q^2$ in the range $x<0.01$ 
consistent with the QCD fit.
Therefore the data were fitted assuming the power behaviour 
$F_2 = c(Q^2) x^{-\lambda(Q^2)}$. The results for the
$\lambda$ and $c$ values are presented in Fig. \ref{lambda} (right).
At $Q^2 < 2$~GeV$^2$ the H1 data were combined with
data of NMC~\cite{nmc} and ZEUS~\cite{zeusbpt}.
We can state that no damping effects of the rise of $F_2$ 
are visible yet at present energies for $Q^2 > 0.85$~GeV$^2$.
For $Q^2 \geq 3.5$~GeV$^2$ and $x < 0.01$,
$F_2$ can be well described
by the very simple parameterisation $F_2 = c x^{-\lambda(Q^2)}$
with $c \approx 0.18$ and $\lambda(Q^2) = a \cdot $ln($Q^2/\Lambda^2$).
At very low $Q^2$ $\lambda$ is approaching 0.08 
which corresponds to the energy dependence of soft hadronic interactions 
$\sigma_{tot} \sim s^{\alpha_P(0)-1} \approx s^{0.08}$~\cite{dola}.

\subsection{High $Q^2$ HERA data}

New high $Q^2$ HERA data were presented by 
M.~Ellerbrock~\cite{elle}, M.~Moritz~\cite{mori} and S.~Grijpink~\cite{grij}.
Both ZEUS and H1 have results from $\sim 16 pb^{-1}$ of $e^-p$ data taken 
in the years 1998-1999 and $\sim 60 pb^{-1}$ of $e^+p$ data taken in the years
1999-2000, both at $\sqrt s = 318 \GeV$. These $e^+p$ data can be combined with
the previously published data at $\sqrt s = 300 \GeV$ to give a total sample
of $\sim 90 pb^{-1}$.
\begin{figure}[ht] 
\begin{center}
\epsfig{file=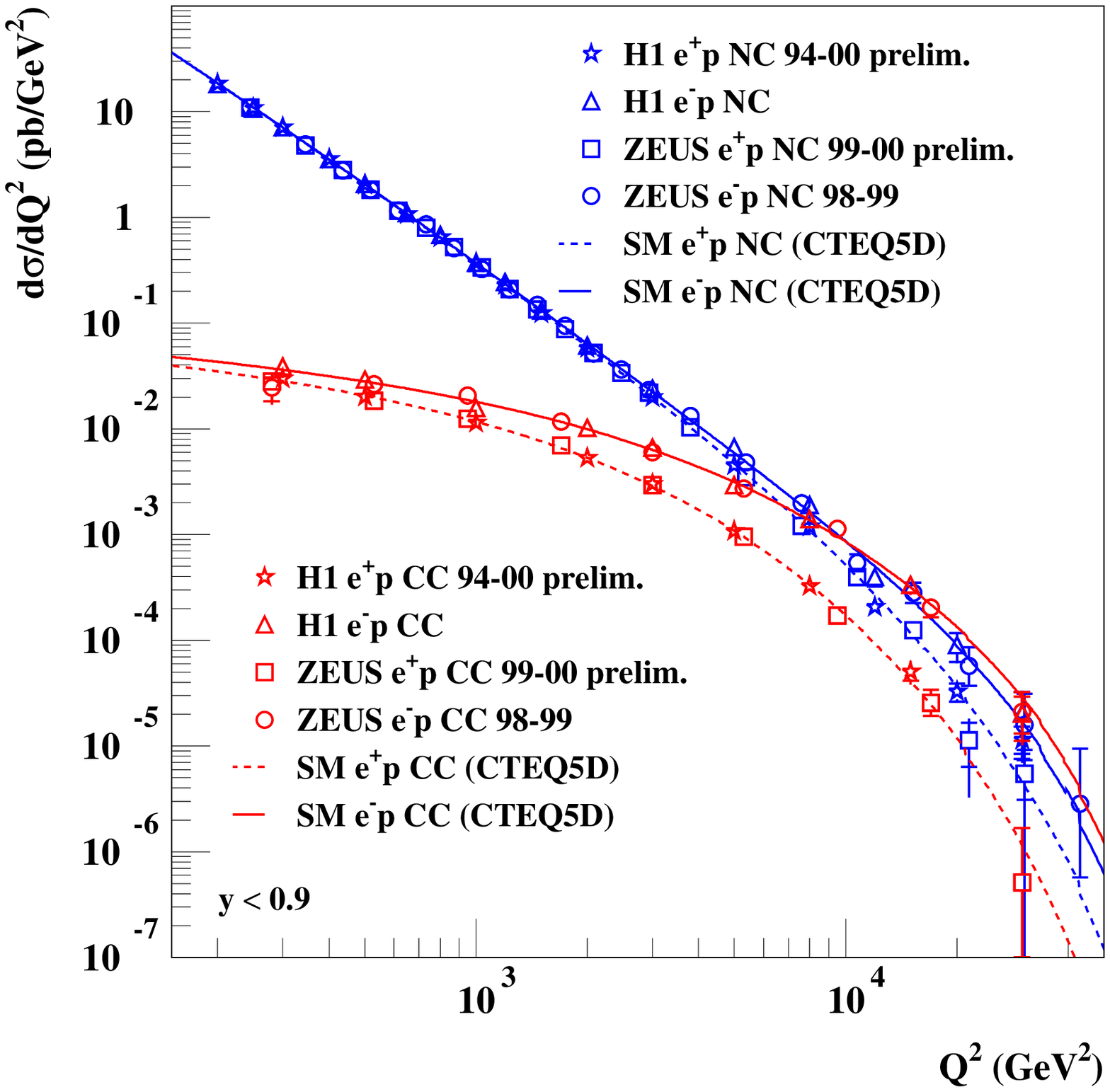,width=6.2cm} 
\epsfig{file=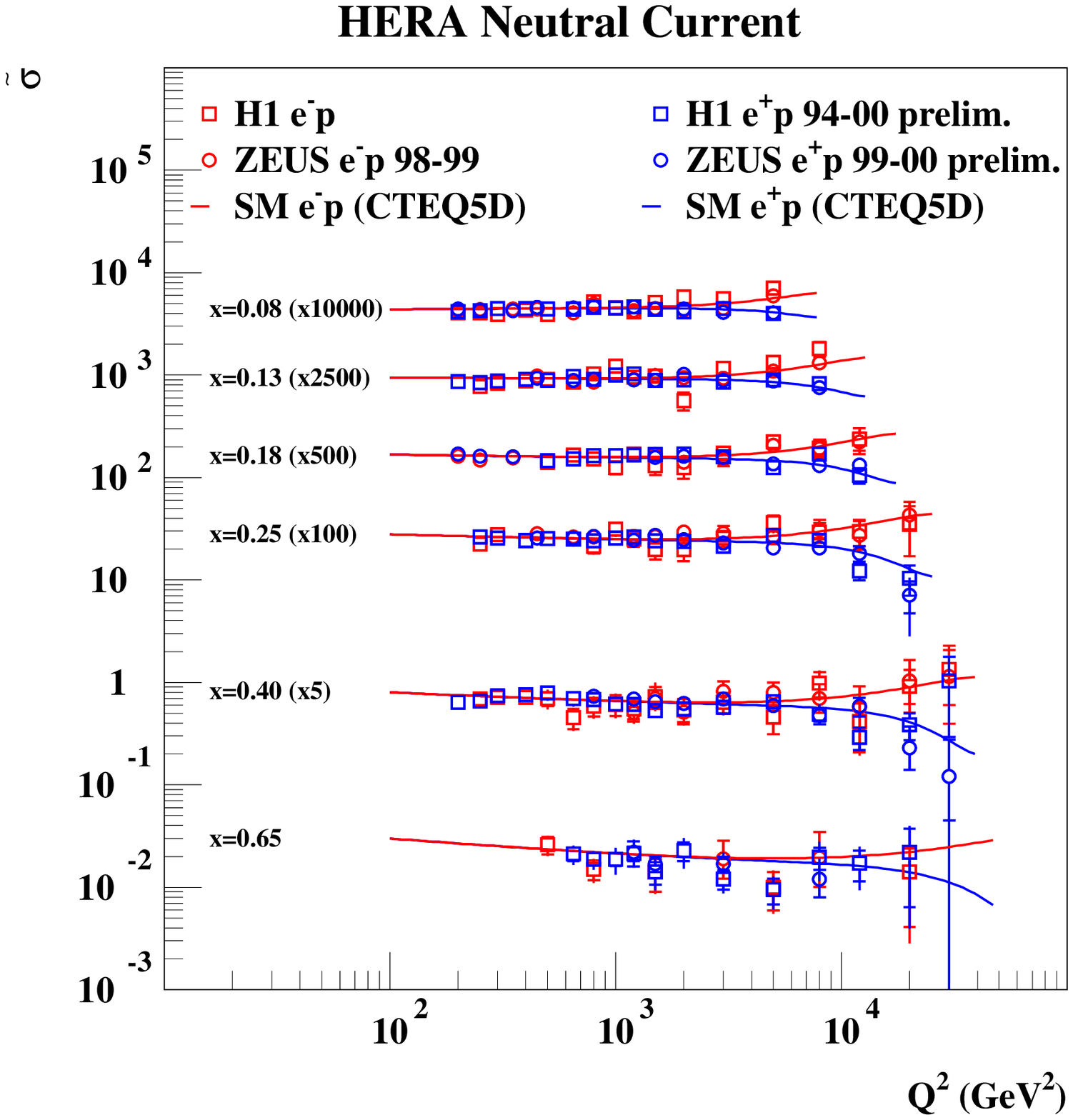,width=6.2cm} 
\caption { \sl Comparison of the ZEUS/H1 data on $e^{\pm}p$ CC and NC scattering
(left). The NC $e^{\pm}$ reduced cross-sections as a function of 
$Q^2$ (right).}
\label {ncccq2}
\end{center}
\end{figure}

The ZEUS and H1 data for neutral (NC) and charged current (CC)
$e^{\pm}p$ scattering are compared in Fig.~\ref{ncccq2}.
There is excellent agreement between the experiments and with the Standard
Model predictions for electroweak unification at high $Q^2$.

\begin{figure}[ht]
\begin{center}
\epsfig{file=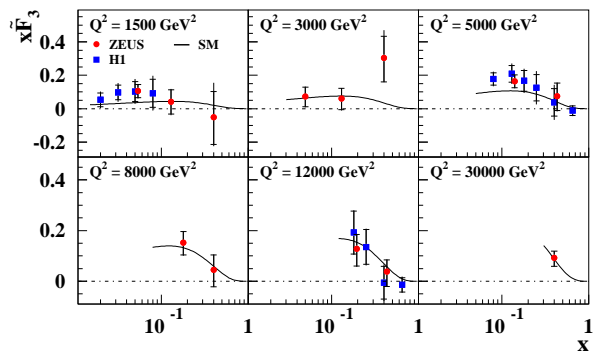,width=0.6\textwidth}
\caption{\sl The structure function $xF_3$ extracted from ZEUS high $Q^2$ 
NC $e^{\pm}p$ data.}
\label{fig:xf3}
\end{center}
\end{figure}

Neglecting the small contribution of $F_L$, the differential cross-section 
 for NC $e^{\pm} p$
scattering is given by
{\small \begin{equation}
\frac {d^2\sigma (e^{\pm}p) } {dxdQ^2} =  \frac {2\pi\alpha^2} {Q^4 x}  
\left[Y_+ F _2(x,Q^2) \mp Y_- xF_3(x,Q^2) \right],
\label{eq:NCxsec}
\end{equation}} 
where $F_2, xF_3$ are expressed in terms of parton distribution functions
(PDFs) as
\begin{equation}
F_2(x,Q^2)= \Sigma_i A_i(Q^2) (xq_i(x,Q^2) + x\bar q_i(x,Q^2))
\end{equation}
and
\begin{equation}
xF_3(x,Q^2)= \Sigma_i B_i(Q^2) (xq_i(x,Q^2) - x\bar q_i(x,Q^2))
\end{equation}
in leading order perturbative QCD.
For unpolarised lepton beams the coefficients $A, B$ are given in terms 
of electroweak couplings~\cite{cddrev}.
The parity violating structure function $xF_3$ is only significant at 
high $Q^2$. Fig~\ref{ncccq2} also shows the difference in the $e^+$ and $e^-$
NC cross-sections due to this $xF_3$ term as a function of $Q^2$. This has 
been used  to extract $xF_3$ and a new measurement from ZEUS is shown in
Fig~\ref{fig:xf3}. With the greater luminosity of HERA-II a precison 
measurement
of this valence structure function will be possible across all $x$. 
Currently the only such precision measurement 
is the $xF_3$ measurement from CCFR $\nu, \bar \nu$ scattering on an Fe target.

\begin{figure}[ht]
\begin{center}
\epsfig{file=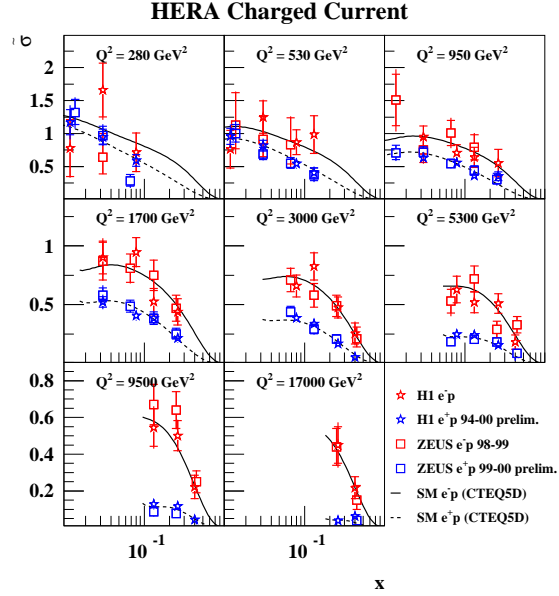,width=0.7\textwidth}
\vspace{-0.5cm}
\caption{\sl CC $e^{\pm}p$ data from ZEUS and H1 }
\label{fig:ccelpo}
\end{center}
\end{figure}

High $Q^2$ CC data can be used to gain information on
the high $x$ valence PDFs, with flavour separation between $u_v$ and $d_v$.
CC scattering involves only the quark flavours which are
appropriate to the charge of the current, so that
the differential cross-section for CC $e^{\pm} p$
scattering with unpolarized beams is given by
\begin{equation}
\frac {d^2\sigma (e^- p) } {dxdQ^2} =  \frac {G_F^2} {2\pi x} \frac {
 M^4_W} { (Q^2 + M_W^2)^2 }\left[  xU(x,Q^2) + (1-y)^2 x\bar D(x,Q^2)\right]
\end{equation}
and
\begin{equation}
\frac {d^2\sigma (e^+ p) } {dxdQ^2} =  \frac {G_F^2} {2\pi x} \frac {
 M^4_W} { (Q^2 + M_W^2)^2 }\left[  x\bar U(x,Q^2) + (1-y)^2 xD(x,Q^2) \right]
\end{equation}
where $U$ stands for $U$-type quarks with charge $+2/3$ and $D$ for $D$-type
quarks with charge $-1/3$. 
Clearly at high $x$ the $e^-p$ cross-section is dominated by the $u_v$ PDF and
the $e^+p$ cros--section by the $d_v$ PDF.
The new high $Q^2$ CC data are shown in Fig~\ref{fig:ccelpo}. Their 
contribution to the precision extraction of PDFs will be discussed in 
Sec.~\ref{sec:qcdres}.

The strong dependence of the CC cross-sections on the $W$ propagator can be 
used to make an extraction of $M_W$ in a space-like process. 
The results from ZEUS and H1 for $e^-$ and
$e^+$ data are given in Table~\ref{tab:mw}. The $e^-$ data 
give the better determinations, both because of the larger cross-section and 
because of the reduced uncertainty from the 
PDFs when the better known $u$ quark
distribution is dominant.
\begin{table}
\begin{center}
\begin{tabular}{|l|c|c|}
\hline
Experiment & beam& $ M_W$\\

\hline
 ZEUS & $e^+$ & $81.4 \pm 2.7(\rm stat) \pm 2.0(\rm sys) \pm 3.0 (\rm PDF)$\\
 H1  & $e^+$&  $80.9 \pm 3.3(\rm stat) \pm 1.7(\rm sys) \pm 3.7 (\rm PDF)$\\ 
 ZEUS & $e^-$ & $80.3 \pm 2.1(\rm stat) \pm 1.2(\rm sys) \pm 1.0 (\rm PDF)$\\
 H1  & $e^-$&  $79.9 \pm 2.2(\rm stat) \pm 0.9(\rm sys) \pm 2.1 (\rm PDF)$\\ 

\hline
\end{tabular}
\caption{Values of $M_W$ extracted from ZEUS and H1 CC data
}
\label{tab:mw}
\end{center}
\end{table}

At HERA-II our ability to measure electroweak parameters will be greatly 
improved both due to increased statistics and due to the polarization
of the beams, as detailed in the contribution of F.~Metlica~\cite{metlica}. 
For example,
 with $1 fb^{-1}$ and $P(e^-) = -0.7$, the error
achievable will be $\Delta M_W \sim 0.055 \GeV$, c.f. the PDG value 
$\Delta M_W \sim 0.049 \GeV$ from time-like processes.  

\section{Recent NLO QCD fits}
\label{sec:qcd}
A large amount of new data has become available during the past couple
of years, in particular the recent measurements of inclusive DIS cross
sections in $ep$ interactions by H1 and ZEUS and the inclusive
high-$E_T$ jet data by D0 and CDF.  The improved precision of the data
led to a new generation of global NLO DGLAP QCD analyses, such as
MRST01~\cite{mrst01} and CTEQ6~\cite{cteq6} presented at the workshop
by R. Thorne~\cite{thorne} and W.K. Tung~\cite{tung}.
B. Reisert~\cite{reisert} and E. Tassi~\cite{tassi} presented QCD fits
performed by H1~\cite{h1lowq2} and ZEUS using their respective data
supplemented by the data from fixed target experiments.  Improved
quality of the fits is achieved due to more precise data as well as
due to a new level of sophistication in the fitting technique
including a full treatment of available experimental correlated
systematic uncertainties.

\begin{figure}[ht]
\begin{center}
\vspace{-2.0cm}
\epsfig{file=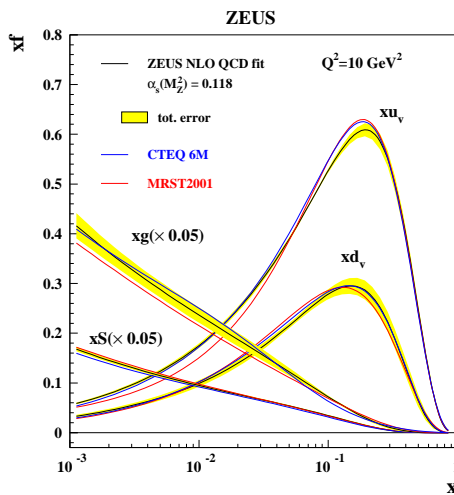,width=0.6\textwidth}
\vspace{-2.0cm}
\caption{\sl Comparison of the ZEUS, MRST2001 and CTEQ6 PDFs. The error band
is that of the ZEUS standard (ZEUS-S) global fit.}
\label{fig:pdferr}
\end{center}
\end{figure}

The MRST01, CTEQ6 and ZEUS PDFs are compared in Fig.~\ref{fig:pdferr},
where the error band illustrated is that from the ZEUS standard (ZEUS-S)
analysis. There is good agreement of all these PDFs within experimental
uncertainties.

\subsection{Uncertainty of parton distributions from QCD fits}
\label{sec:qcderr}
\begin{figure}[ht] 
\begin{center}
\epsfig{file=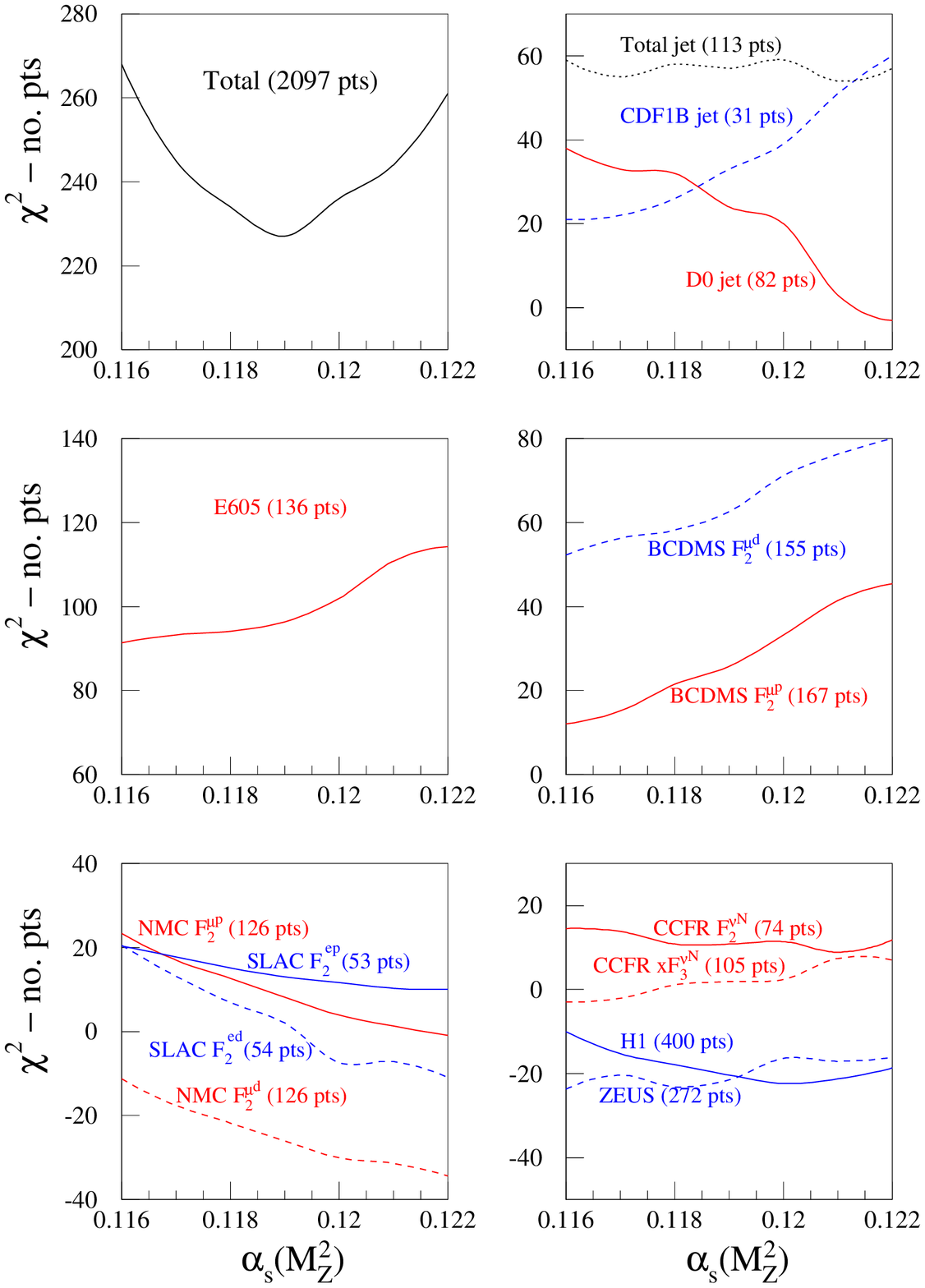,width=6.2cm} 
\epsfig{file=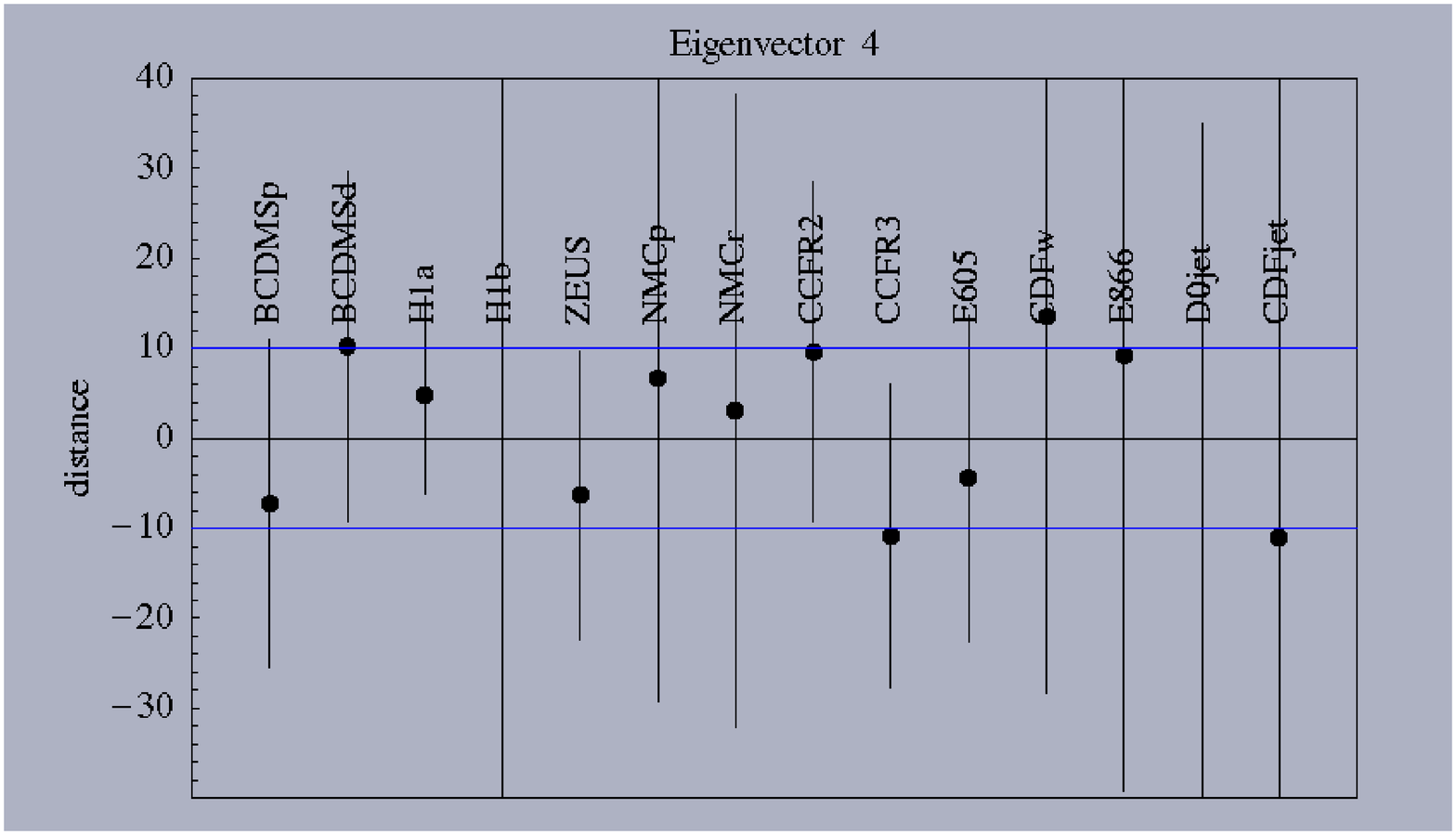,width=6.2cm} 
\caption { \sl 
Partial $\chi^2$ for data sets in the MRST01 fit 
as function of $\alpha_s(M_Z^2)$ (left).
Distance along a parameter combination (eigenvalue 4, CTEQ6)
from the $\chi^2$-minimum of an individual data set 
to the global minimum (right).
In the neighborhood of the global minimum a distance of 1
corresponds to $\Delta \chi_{global}^2 \approx 1$.
}
\label {chi2}
\end{center}
\end{figure}

A new feature of the recent QCD analyses is a systematic and pragmatic
treatment of the uncertainties of the parton distribution functions 
and their physical predictions.
One of the problems of the uncertainty estimation 
for fits with many data sets is related to
a certain degree of inconsistency of the latter.
Usually the one sigma error of a parameter in a fit is 
determined by variation of $\chi^2$ by one unit from the minimum.
Very often, however, this rule becomes unrealistic.
This is demonstrated in Fig.~\ref{chi2}, where the distances
from $\chi^2$-minima of individual data sets 
to the global minimum by far exceeds 
the range allowed by the $\Delta \chi^2 =1$ criterion.
It is not possible to simply drop ``inconsistent'' data sets, 
as then the partons in some regions would lose important constraints.
On the other hand the level of ``inconsistency'' should 
be reflected in the uncertainties of the PDFs.
This can by achieved by modification of the $\chi^2$ tolerance criterion to
$\Delta \chi^2 =T^2$~\cite{cteq6,mrst01,durham,mydur}
where $T$ stands for a tolerance which should be
estimated from the level of (in)consistency of the data sets
used in each particular QCD fit.
In the CTEQ6 fit the tolerance was taken to be 10
($\Delta \chi^2 =100$), as shown by the
horizontal lines in Fig.~\ref{chi2} (right). 
The choices for $T^2$ in the QCD fits
are listed in Table~\ref{tab:alf} and range from 1 to 100.

\begin{table}
\begin{center}
\begin{tabular}{|l|c|cc|}
\hline
  CTEQ6 & $\Delta \chi^2 = 100$ & $\alpha_s(M_Z^2)$=& $0.1165
  \pm 0.0065(exp) $ \\
  ZEUS  & $\Delta \chi_{eff}^2 = 50$ & $\alpha_s(M_Z^2)=$&$ 0.1166
  \pm 0.0049(exp) $ \\ 
  & & & $\pm 0.0018(model) \pm 0.004(theory) $ \\
  MRST01 & $\Delta \chi^2 = 20$ & $\alpha_s(M_Z^2)=$&$ 0.1190
  \pm 0.002(exp) \pm 0.003(theory)  $ \\
  H1    & $\Delta \chi^2 = 1$ & $\alpha_s(M_Z^2)=$& $0.115 \pm 0.0017(exp)$ \\
  & & & $ ^{+~~0.0009}_{-~~0.0005}~(model) \pm 0.005(theory)$ \\
\hline
\end{tabular} 
\caption{\sl Values of $\alpha_s(M_Z^2)$ and its error from different NLO QCD 
fits with different error tolerances.}
\label{tab:alf}
\end{center}
\end{table}

\begin{figure}[ht] 
\begin{center}
\epsfig{file=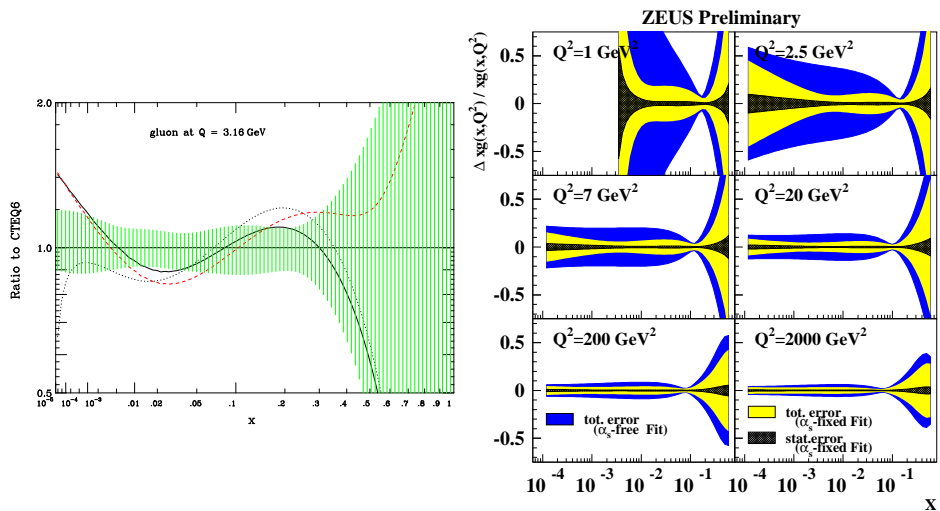,width=6.0cm}
\epsfig{file=pdf_xq2.color.2.epsi,width=6.3cm} 
\caption { \sl 
Relative gluon uncertainties as determined in the CTEQ6 fit (left)
and the ZEUS-S QCD fit (right).
}
\label {errband}
\end{center}
\end{figure}

The values of the strong coupling constant $\alpha_s(M_Z^2)$ obtained
in the fits are also given in Table~\ref{tab:alf}. They are 
remarkably consistent. However, the estimates of the experimental uncertainties
on $\alpha_s(M_Z^2)$  are different due to different
judgements on the $\Delta \chi^2$ criterion.  This is not a
contradiction, all choices are legitimate and reflect different
emphases in the fits.  For example, H1~\cite{h1lowq2} uses the
canonical $\Delta \chi^2 =1$ after careful consistency checks of the
two data sets (H1 and BCDMS $\mu p$) used in the fit.  The relative
uncertainty bands for the gluon distribution obtained in the CTEQ6 and
ZEUS-S fits are shown in Fig.~\ref{errband}. They illustrate a reasonable 
consistency of judgement on the experimental errors of the gluon PDF.

\begin{figure}[ht] 
\begin{center}
\epsfig{file=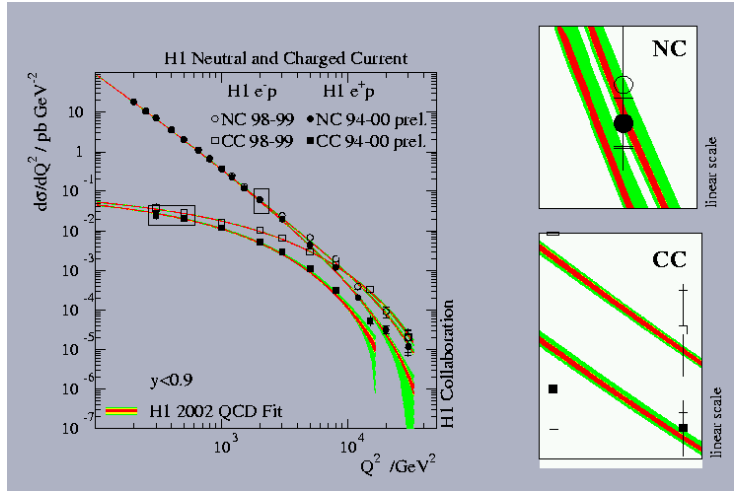,width=6.5cm,angle=-90} 
\caption { \sl 
Single differential cross sections $d \sigma/d Q^2 $
for NC and CC processes 
in the $e^{\pm}p$ interactions (left).
Grey error bands are envelopes of the QCD fit results
corresponding to variations of parametric forms. 
The boxes in the left figure for NC and CC at moderate $Q^2$ 
are zoomed and shown on the right.
}
\label {nccc}
\end{center}
\end{figure}

Thus, there are reasonable approaches to how to treat 
experimental statistical and systematical errors, how to take into
account model uncertainties such as charm or bottom masses, and how to
account for incompatibilities of data sets.  It is not so easy to estimate
theoretical uncertainties, this is explored further in~\cite{thorne} and
in Sec.~\ref{sec:coll}. The model uncertainty which comes from
the choice of parametric forms
for the PDFs at the input scale also merits further investigation.  
The latter issue was studied by
CTEQ~\cite{cteq6} and H1.  B.~Reisert~\cite{reisert} presented an
investigation of the parameter space using general forms of MRST type
parameterisations $ x \cdot PDF = a x^b (1-x)^c (1 + d \sqrt x + e x)
$ for the PDFs of the gluon, quarks and anti-quarks at the input
scale.  Starting with the parameters $a, b, c$ for each PDF, an
additional parameter was considered only when its introduction
improved $\chi^2$ by more than one unit.  The uncertainty envelope in
this study was defined as an overlap of the experimental and model
error bands of fits with $\chi^2 < \chi_{best}^2 +\sqrt {2 N_{dof}}$.
Here $\chi_{best}^2$ corresponds to the fit with the best
$\chi^2$. This criterion is constructed by analogy with the
statistical error of $\chi^2$.  It is somewhat arbitrary and used here
only provisionally.  The uncertainty bands in Fig. \ref{nccc} show
that an assumption on the parametric form for the PDFs at an input
scale of $Q_0^2 = 4$~GeV$^2$ influences the predictions of the QCD fit
even at moderate $Q^2$ in the case of highly integrated observables
like $d \sigma/d Q^2 $.

\subsection{Results from QCD fits}
\label{sec:qcdres}
\begin{figure}[ht] 
\begin{center}
\epsfig{file=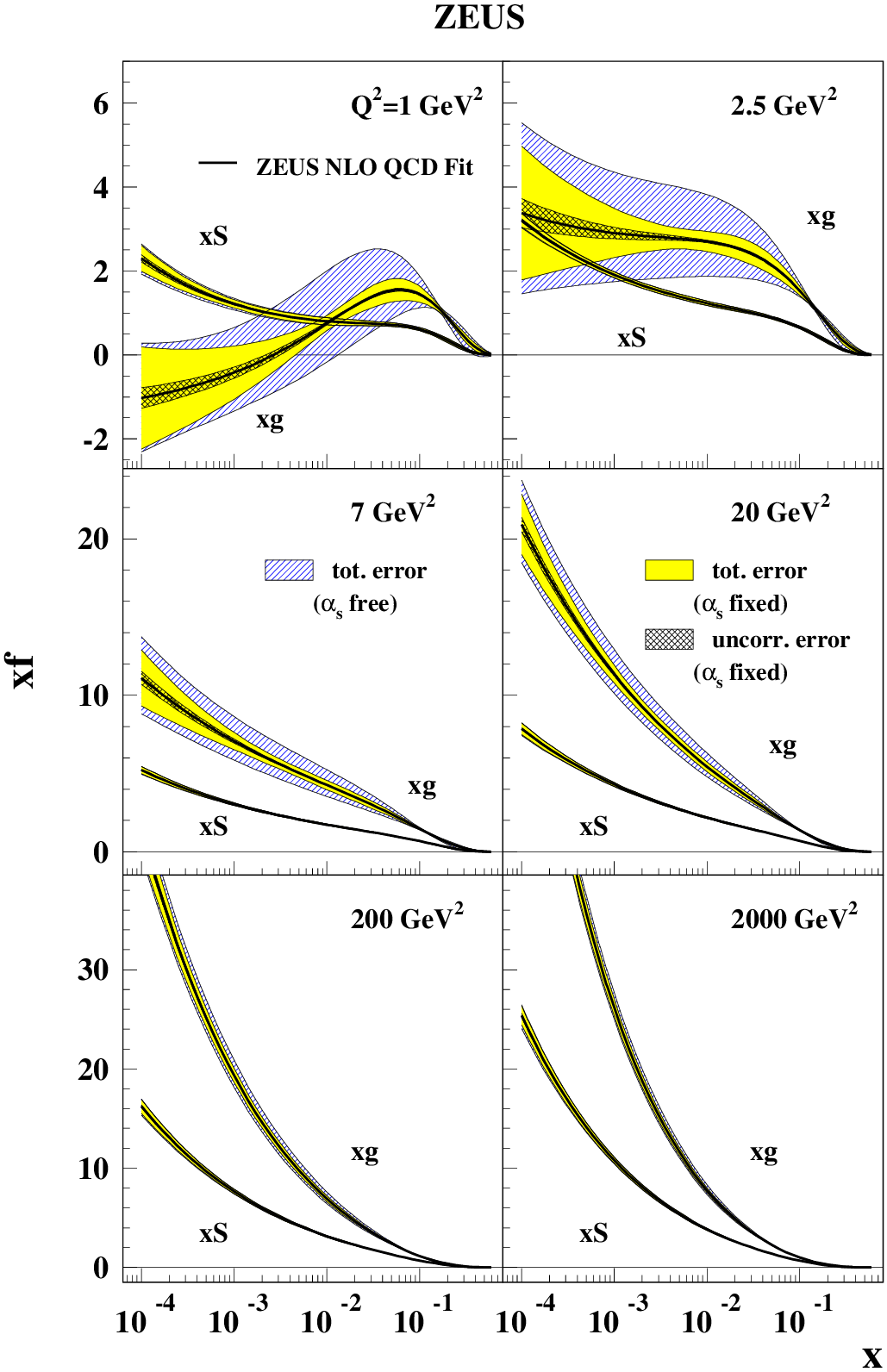,width=6.2cm} 
\epsfig{file=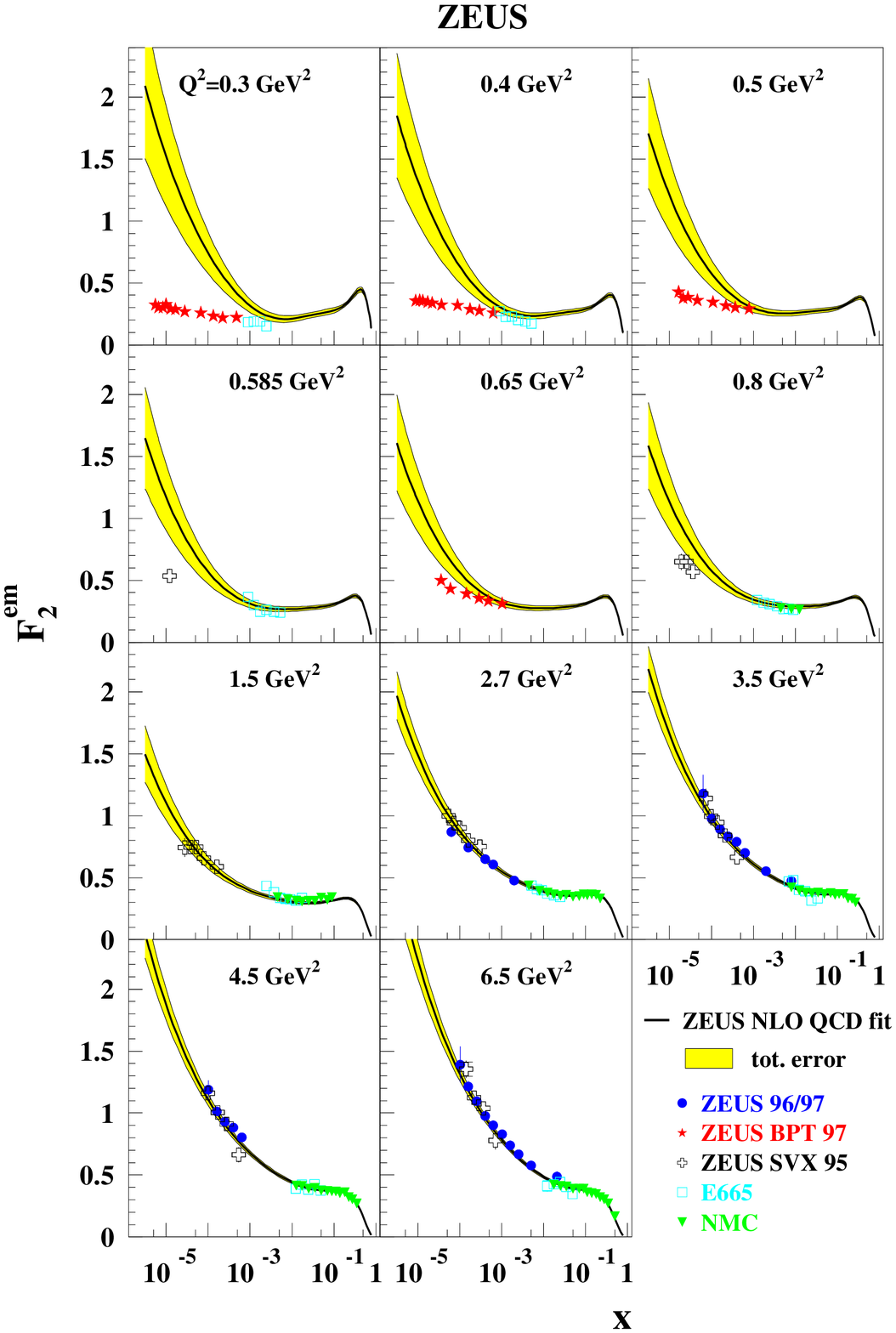,width=6.2cm} 
\vspace{0.0cm}
\caption { \sl Gluon and sea distributions for the ZEUS-S fit
(left). $F_2$ data at very low $Q^2$ compared to the ZEUS-S fit.
(right).}
\label {fig:glusea}
\end{center}
\end{figure}

The HERA data are crucial for determining
the low $x$ sea and gluon shapes. The ZEUS-S sea and gluon distributions
are compared in Fig~\ref{fig:glusea}~\cite{tassi}. 
The gluon density is much larger than 
the sea density for $Q^2 > 5$~GeV$^2$, but for lower $Q^2$ the sea density
continues to rise at low $x$ (consistent with the rise in $F_2$ down to low
$Q^2$ mentioned in Sec~\ref{sec:lowq2}), whereas the gluon density is 
suppressed. This could be a signal that the conventional DGLAP formulation
of NLO-QCD is inadequate in this region.
Fig.~\ref{fig:glusea} also shows data at very low $Q^2$ compared to the
ZEUS-S fit. Such a fit clearly fails for $Q^2 \leq 0.65$~GeV$^2$, even when 
the conservative error bands on the fit are considered. 

For this generation of PDFs the data which most strongly
determine the valence distributions are the fixed target data.
However new high $Q^2$ HERA data can put some constraint on the
valence distributions and this is important because these data do not
suffer from the uncertainties associated with nuclear target
corrections.

\begin{figure}[ht] 
\begin{center}
\epsfig{file=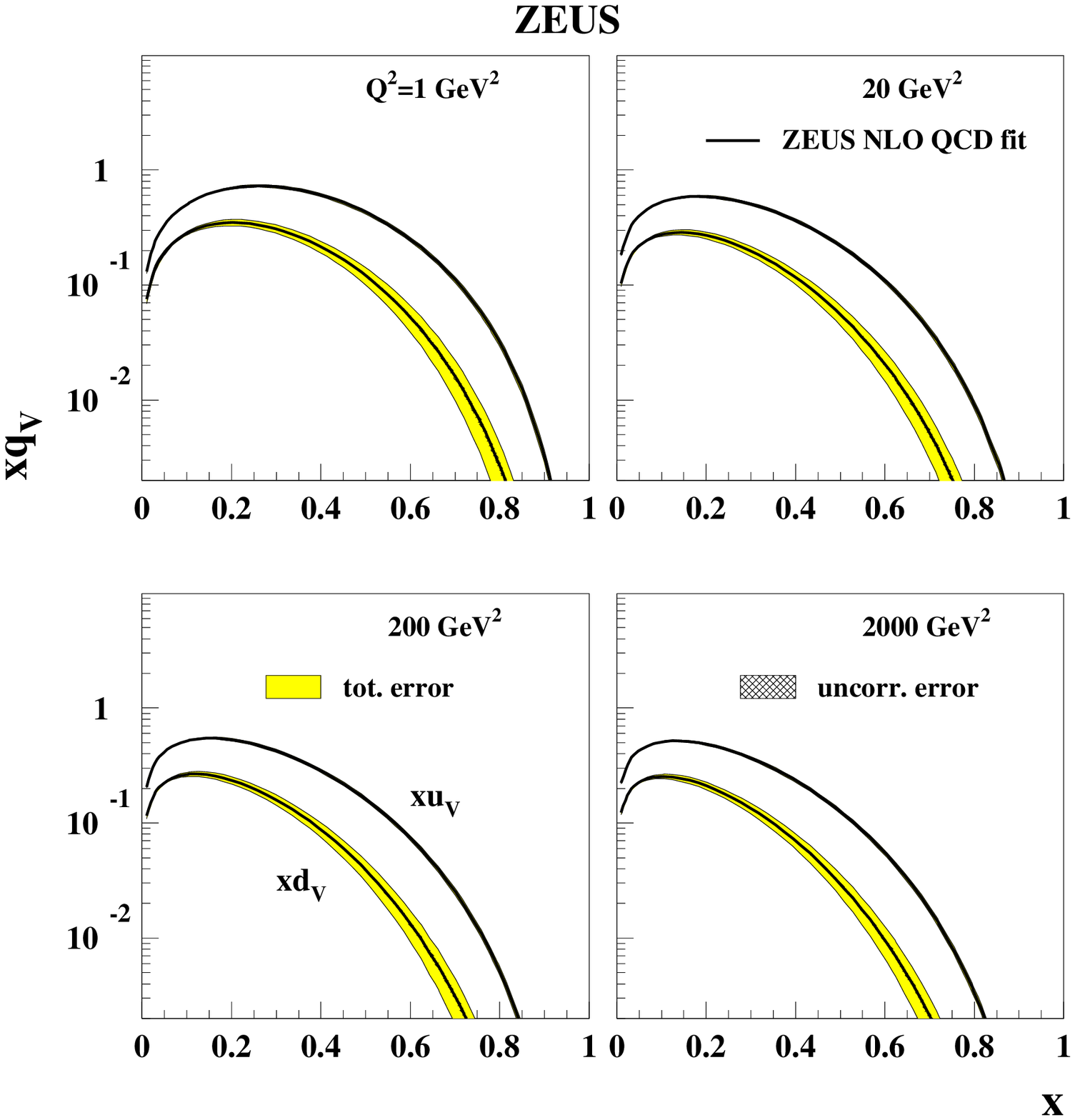,width=6.2cm} 
\epsfig{file=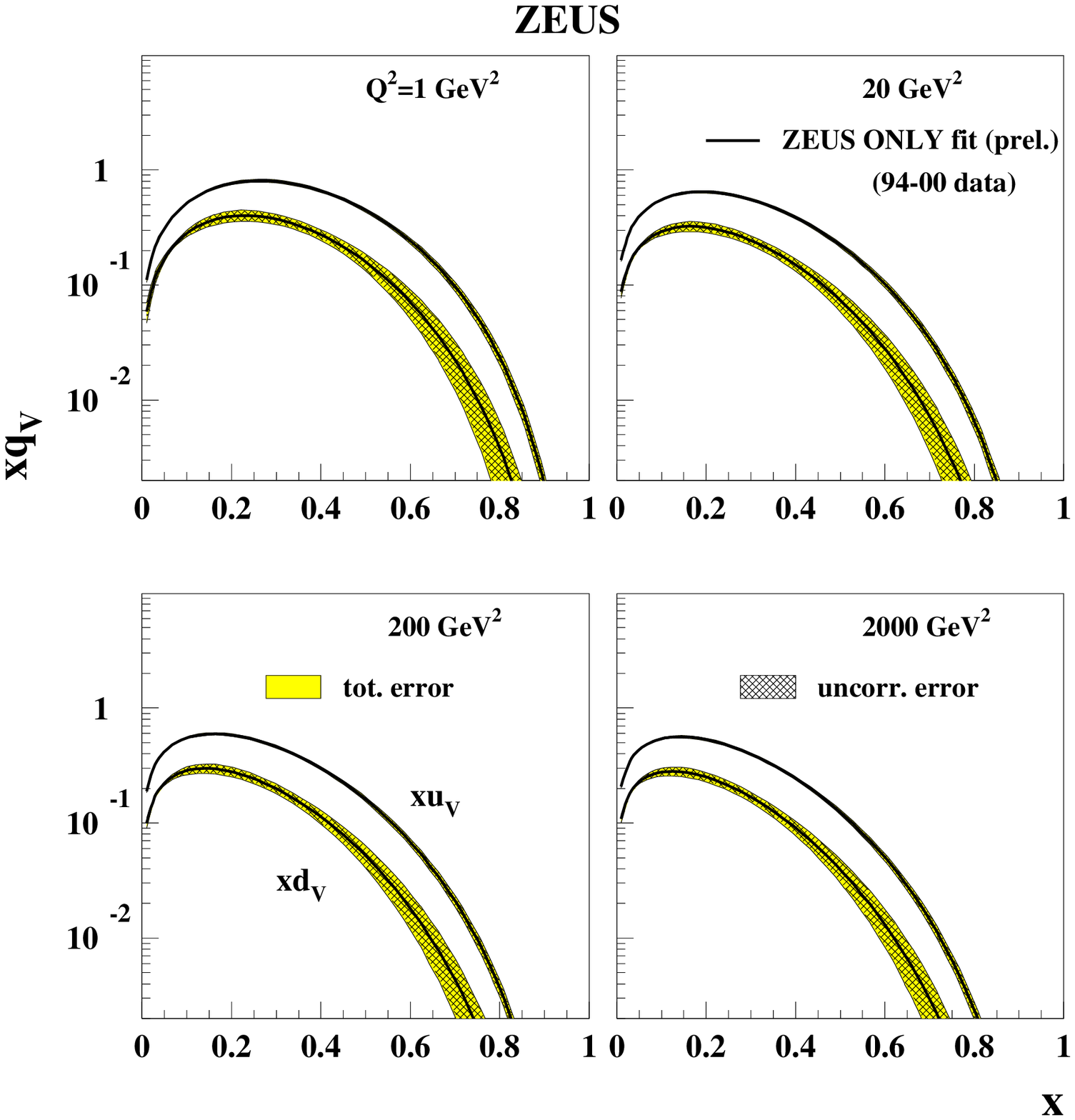,width=6.2cm} 
\vspace{-3.0cm}
\caption { \sl Valence distributions for the ZEUS-S fit
(left). Valence distributions for the ZEUS-O fit
(right).}
\label {valence}
\end{center}
\end{figure}

Both the NC and CC high-$Q^2$ data are very well described by the
global PDF fits. ZEUS has also made a special fit to ZEUS data alone
(ZEUS-O) including the new $e^-p$ 98/99 and the preliminary $e^+p$
99/00 high $Q^2$ data~\cite{tassi}. In this fit these additional data sets were
used instead of the fixed-target data to constrain the valence
distributions. Fig.~\ref{valence} compares the valence distributions
from the ZEUS-S global fit to those for the ZEUS-O fit. The level of
precision of the ZEUS-O fit is approaching that of the global fit and
its precision is statistics limited rather than systematics limited,
so that improvement can be expected with higher luminosity HERA-II
data. The systematic precision of high-$x$ ($x > 0.7$) measurements at
HERA-II can also be improved further as explored in the contribution of
M.~Helbich~\cite{helb}.

There are further advantages to using HERA data alone.
In the ZEUS-O fit, the high-$x$ $d$-valence distribution is determined 
by the high-$Q^2$ $e^+p$ CC data.
In contrast in the global fits it is strongly
determined by the NMC $F_2^D/F^p_2$ data.
It has been suggested that such measurements are 
subject to significant uncertainty from deuteron binding
corrections~\cite{bodek}. The ZEUS-O extraction does not suffer
this uncertainty.  

The ZEUS-O fit was made using the same form of parton parametrization
as the ZEUS-S global fit. For the global fits, parametrization dependence
is not severe since, for example, 
the valence shapes are strongly constrained across all $x$ by the
CCFR $xF_3$ data, and the $\bar d - \bar u$ distribution is constrained by
$D,p$ target data. However, if HERA data alone are used, then these constraints
are lost and parametrization dependence can be significant, as discussed
in Sec~\ref{sec:qcderr}~\cite{reisert}. At HERA-II the precision 
measurement
of $xF_3$ across all $x$ should considerably reduce this uncertainty, and
eliminate the uncertainty from heavy target corrections which is unavoidable in
the CCFR $xF_3$ measurement.

\section{$F_2^{c\bar c}$, $F_2^\gamma$}

Results relevant to the charm structure function and the photon structure
function were extensively covered in the Hadron Final States Working Group
and will not be discussed again in detail here.
Results from HERA on the charmed 
structure function were reviewed for the Structure Function Working Group by
O.~Behnke~\cite{behnke}. 
M.~Prybycien~\cite{pryb} reviewed the status of photon structure function
measurements at LEP.A.~ De~Roeck~\cite{roeck} presented new OPAL data from 
two-photon processes and S.~Maxfield~\cite{maxfield} reviewed measurements 
of real and virtual photon structure. There was also some progress in
developing new PDF sets for the photon, accounting for modern data and
correct heavy quark treatment, as presented by 
S.~Albino~\cite{albino} and P.~Jankowski~\cite{Janowski}.

\section{Theory}
\label{sec:theory}
At this workshop the main emphasis of the theoretical contributions was 
the different ways in which one can calculate structure functions and
the regions of applicability of these different approaches. Essentially 
there were four alternative methods which were outlined, all of which have 
seen significant progress, or at least new results. These are:

\medskip

\noindent 1. Lattice QCD.

\noindent 2. Saturation type effects/colour glass condensates.

\noindent 3. $k_T$-factorization.

\noindent 4. Collinear Factorization.

\medskip

There were also some other talks which do not fall into these general 
categories. D.~Haidt presented a consistency check for DGLAP evolution
\cite{haidt}, 
examining the partons extracted from the measured values of $F_2$ and 
$dF_2/d \ln Q^2$ and checking that these are consistent with the evolution 
equations. A discrepancy is found at low $x$ and $Q^2$. T.~Lastovicka
demonstrated  that a good fit to structure functions may be obtained using 
a parameterization determined by assuming a self similar structure, i.e. 
using the fractal dimensions for the structure functions \cite{lastovicka}.
A.~Kotikov 
presented a fit to high $x$ data using cuts determined by the region 
of large systematic errors extracting, for example,  
$\alpha_S(M_Z^2) = 0.1174\pm
0.0007({\rm stat})\pm 0.0019 ({\rm sys}) \pm 0.0010({\rm norm})$ 
from a nonsinglet fit \cite{kotikov}.
D.~Timashkov also presented an analytic formula for structure functions  
for all $x$ and $Q^2$ based on expressions in the limiting cases $Q^2 \to 0$,
$x \to 1$ and $x \to 0$ \cite{timashkov}. However, the summary is based on 
the above four alternative procedures.

\subsection{Lattice QCD} 

There has been significant progress in this area, and we were given a 
summary by S.~Capitani \cite{capitani}. 
It is not possible to compute structure functions directly on the lattice 
because the parton distributions are defined on the light cone, while lattice 
simulations are done in Euclidean space. However, one can use the Operator
Product Expansion and calculate moments.    
The main effort has been in the calculation of 1st, 2nd and 3rd 
moments of nonsinglet distributions, both for unpolarized and polarized 
structure functions. A reason for only calculating nonsinglet quantities 
is due to the difficulty in computing disconnected diagrams (i.e., connected 
only by gluon lines) due to the expense in computer time. Nonsinglet 
quantities are insensitive to such diagrams.
 
One of the main improvements has been 
the first calculations without using the quenching approximation. This has 
shown that, for nonsinglet quantities at least, the quenched 
approximation is indeed very good.  
There have also been improvements in the perturbative renormalization 
factors required to 
translate the results on the lattice to a particular continuum 
renormalization scheme (e.g ${\overline {\rm MS}}$). 
In order to obtain the final 
results on the lattice it is ultimately necessary to perform chiral and 
continuum extrapolations, using a fit formula $A+ B m_{\pi}^2 + ca^2$,  
due to the finite lattice spacing $a$ and to the fact that one currently
has a pion with mass $m_{\pi} \sim 500\MeV$. This appears to be well under 
control, but the results are disappointing -- for the first moment of the 
$u-d$ distribution they find $0.30 \pm 0.03$ where the standard distributions
give $0.23\pm 0.02$. The results for polarized distributions are more in 
agreement with experiment.

It is thought that this discrepancy is due to the finite size of the lattice 
missing the effects of the pion cloud. From chiral perturbation theory one 
obtain terms $\sim m_{\pi}^2\ln m_{\pi}^2$. Introducing an additional 
term in the extrapolation formula of the form $m_{\pi}^2\ln( 
m_{\pi}^2/\Lambda^2)$ can solve this problem, but only for $\Lambda$ a free
parameter $\sim 300-700\MeV$ for various processes, destroying any
predictive power. One needs lattices such that 
$m_{\pi} < 250\MeV$, which may be possible with computers within a couple 
of years. It would also be desirable to investigate the pion cloud effects 
by doing simulations with larger physical volumes. Finally, preliminary 
investigates of higher twist moments have been performed, giving results 
which are surprisingly small.

\subsection{Saturation type effects/colour glass condensates}

There was a lot of emphasis on the region of small $x$ and low $Q^2$ 
where the gluon density is expected to saturate. Recently
a great deal of interest has focused on the 
GBW saturation model \cite{gbw}. 
In this one factorizes deep inelastic scattering into the fluctuation of the 
virtual photon into a dipole pair and the dipole-proton cross-section. 
The former is calculable at LO and the latter is modelled in the form 
$$
\hat \sigma(x,r) = \sigma_0 [1-\exp(-r^2/4R_0^2(x))]
$$
where $r$ is the dipole size and $R_0^2(x)=(x/x_0)^{\lambda}\GeV^{-2}$. 
This then saturates at large $r$ and low $x$ and predicts  
geometric scaling, i.e that the structure functions are functions of 
$Q^2R_0^2(x)$ alone. For $\lambda \sim 0.28$ this model could be made to fit
data well, and the procedure could then predict the total diffractive 
cross-section. We had a presentation from N.~Timneanu on an extension of 
this type of approach to two photon physics \cite{timneanu}. Using a number of
sensible extensions of the dipole-proton cross-section to model the 
dipole-dipole cross-section he demonstrated that a good fit to existing
data on photon-photon cross-sections is obtained for both real and virtual 
photons. However, this data is not very precise, and in particular, can not 
distinguish between the different extensions.  

\begin{figure}[htbp]
\begin{center}
\includegraphics[width=9cm]{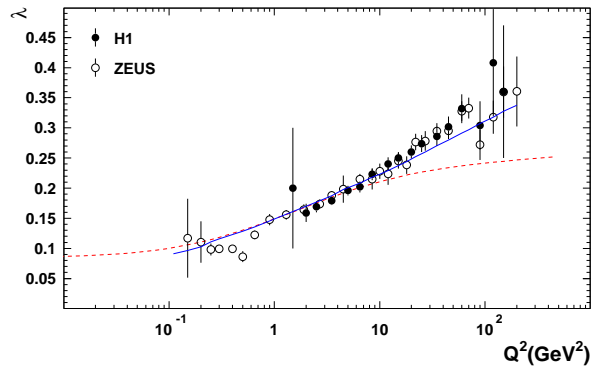}
\end{center}
\vspace{-0.6cm}
\caption{The effective slope $\lambda$ as a function of $Q^2$ - the original
saturation model (dashed line) and the improved model (dotted line).}
\label{gbwfig}
\end{figure}

However, the new more accurate HERA data on proton structure functions 
presented at this 
workshop is very precise and forces the simple saturation model 
to be modified, e.g.,
the effective power $\lambda$ in $F_2(x,Q^2)\sim x^{-\lambda}$ is accurately
measured to run with $Q^2$ (see Sec.~\ref{sec:lowq2} and Fig.~\ref{lambda}) 
whereas it saturates at moderate $Q^2$ in the 
simplistic saturation model. In order to rectify this it is necessary to 
include DGLAP evolution in the model \cite{gbwnew}, 
replacing the above $\hat \sigma(x,r)$ by
$$
\hat \sigma(x,r) = \sigma_0 \biggl[1-\exp\biggl(\frac{-\pi^2r^2\alpha_s
(\mu^2) xg(x,\mu^2)}{3\sigma_0}\biggr)\biggr],
$$
where $\mu^2$ is parameterized in terms of $r^2$. This can improve the fit 
considerably, as shown in Fig.~\ref{gbwfig}. 
However, it moves one away from the 
original principles somewhat, and geometric scaling is violated.  
There are two alternative formulations in
\cite{gbwnew} (somewhere between the two is likely to be most realistic), 
both moving the saturation scale to lower $x$ and $Q^2$. Including 
charm quarks rather than only 3 light flavours in the original model
(hardly optional since charm contributes over $30\%$ of the structure 
function in some regions) also moved the saturation scale down 
an order of magnitude in $x$. If this is also true for the modified model 
it implies that for $Q^2=1\GeV^2$ saturation sets
in only for $x < 0.00001$ in structure functions.

\vspace{-1cm}

We also had some other talks on saturation 
in QCD, discussing the solutions to the nonlinear equations describing
QCD at high parton density and adding support to the type of models 
considered above. We had a review of this subject by L.~McLerran 
\cite{mclerran} (which will be discussed in the summary of the diffractive 
session). We also had a presentation by A.~Freund which outlined some   
specific points \cite{freund}. He told us that  ``Colour 
Glass Condensate'' simply means that in QCD (colour) at high energies, fields 
slowly evolve relative to natural scales (glass) and the phase-space density
saturates (condensate). Hence, at high energies, but at
scales high enough that perturbative QCD applies, 
one can write a renormalization group 
equation in $x$ in terms of suitable variables \cite{weigert}. 
This results in a 
Fokker-Plank equation which is nonlinear, but which one can use known 
techniques to solve. Implementing a single input condition one obtains a 
solution (for fixed $\alpha_S$) of the general form 
$F_2(x,Q^2) = (x/x_0)^{2\lambda} 
f\bigl(x_0, \bigl(\bigl(\frac{x}{x_0}\bigr)^{\lambda}\frac{Q}{Q_0}
\bigr)^2\bigr)$ where $\lambda=0.18$, 
which fits the data well and is similar to geometric scaling. 
This has also been applied to nuclear structure functions with successful 
results, and where it is found that the scaling variable is 
$Q^2(x/x_0)^{2\lambda}A^{-\delta}$, where $\delta =0.1$ rather than the 
naively expected $1/3$. Hence, the first principle solutions in the 
colour-glass-condensate approach support the saturation models, 
implying a similar type of scaling. Corrections to this scaling have been 
estimated. However, there are further improvements to be made, e.g. a full 
treatment of running coupling, and it is intriguing that 
geometric scaling should work 
so well when the charm quark contribution is ignored, despite the fact that
it contributes a great deal to the structure function, and 
should lead to violation of any geometric type scaling.

\subsection{$k_T$ factorization} 

This should be applicable at high energies at scales where perturbation 
theory holds but high density effects are minimal. 
There have been a number of improvements in this field. One of these is in 
the Monte Carlos based on $k_T$-factorization. H.~Jung showed that 
there has been significant progress in correcting previous shortcomings in 
the CASCADE Monte Carlo \cite{cascade}, 
both in the treatment of the scale in the running of the coupling
and in the inclusion of the non-singular terms in the ${\cal O}(\alpha_S)$
gluon-gluon splitting function \cite{jung}. 
Explicitly, in the original version the
splitting function had the form 
$
P = \frac{\bar \alpha_S}{1-z} + \frac{\bar \alpha_S}{z}\Delta_{ns}
$
where $\Delta_{ns}$ is the non-Sudakov form factor. This misses the 
non-singular terms $\bar \alpha_S(-2 +z(1-z))$ in the LO splitting function.
Since this is a negative contribution its omission leads to a bigger gluon 
at high and moderate $x$.   
This has lead to the modification  
$$
P =\bar \alpha_S\biggl(\frac{z}{1-z}+Bz(1-z)\biggr)+ 
\bar\alpha_S\biggl(\frac{1-z}{z} + (1-B)z(1-z)\biggr)\Delta_{ns}
$$
where $B\sim 0.5$, as well as changes to the form factors. 
Unfortunately, although these 
modifications, or something similar, are 
necessary, the decreased positive contribution 
to the evolution at moderate $x$ actually leads
to worse agreement with data for forward jets, and Tevatron $b$ production, 
with the data showing an excess in both cases. There was also an alternative 
approach to $k_T$-factorization based Monte Carlos presented 
by G.~Miu \cite{miu}. This is based on the Linked-Dipole-Chain model,
and differs mainly in the manner in which partons are separated into
initial and final state emissions.
The resulting integrated gluon distribution obtained from fits to $F_2$ agree
well with standard distributions. While this need not be the case at small
$x$, where $k_T$-factorization and collinear factorization may well differ, it
must be the case at higher $x$, where a correctly modified Monte Carlo
should not significantly alter the conventional results. 
 
A.~Stasto talked on solutions to the LO BFKL equation with running 
coupling \cite{stasto}. 
She argued that if one calculates the purely perturbative contribution to the 
high-energy gluon Green's function one obtains an expansion in $\beta_0
\bar\alpha_S^2 Y$ which is reasonably well-behaved  as long as $\beta_0
\bar\alpha_S^2 Y \leq 0.1$ (beyond this the series diverges), and explicit 
results are known for this series \cite{stasto1}. 
It was also demonstrated that the 
transition to the nonperturbative region is a sudden tunneling-like effect,
rather than due to diffusion as is generally assumed \cite{stasto2}. 
The regime where the alternative
methods of breakdown occur was compared and found to be similar for 
$\bar\alpha_S \leq 0.1$ but the perturbative expansion having a larger range 
of applicability for lower $\bar \alpha_S$ (or in the formal limit of
small $\beta_0$). However, $\beta_0 \bar\alpha_S^2 Y \leq 0.1$ is not in
practice a very wide range, and it is unclear if purely
perturbative calculations of high-energy scattering are really possible in 
a quantitative sense, though higher order corrections may help matters.    

\begin{figure}[htbp]
\begin{center}
\includegraphics[width=9cm]{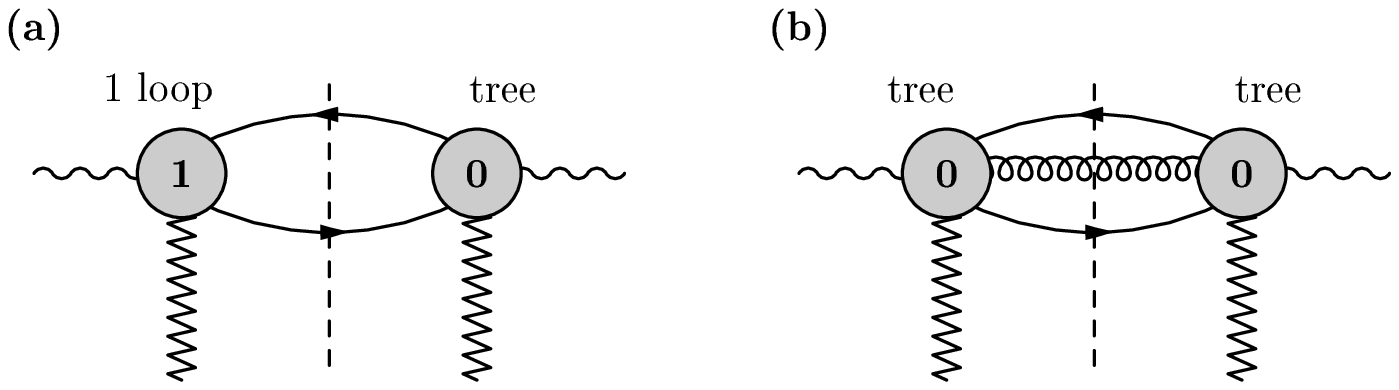}
\end{center}
\vspace{-0.6cm}
\caption{Contributions to the $\gamma^{\star}$ impact factor.}
\label{steffig}
\end{figure}

S.~Gieseke presented an update of the present status of the calculation 
of NLO impact factors in the BFKL framework \cite{gieseke}. This consists of 
two different contributions - the one-loop virtual corrections to the 
quark box diagrams and the contributions with an additional gluon in the 
intermediate state, as illustrated in Fig.~\ref{steffig}. 
The calculation of the vertex
diagrams in each of these two cases is now complete, and moreover, it has been 
proven that the infrared divergences due to the two separate contributions 
cancel each other in the appropriate manner. However, it still remains to 
perform the integrals over phase space to obtain the final result. Once this
is done we will finally be in a position where NLO calculations of physical 
processes can be made with the $k_T$-factorization framework for the first 
time.

\vspace{-0.5cm}

\subsection{Collinear factorization} 
\label{sec:coll}
This is what we normally think of as standard perturbative QCD, which we use
assuming high accuracy at large scales and uncertain $x$, usually at NLO.  
Within the framework of NLO QCD, S.~Kretzer gave a presentation of various 
issues in heavy-flavour production \cite{kretzer}: a 
prescription for fully differential charm production at NLO for charm 
production in DIS \cite{kretzerolness}, 
which will be important in determining the strange quark 
distribution with more accuracy at NuTeV; the NLO and mass corrections to
the DIS contribution to $\nu_{\tau}N \to \tau X$; 
and a modification
of the ACOT prescription for inclusive charm production in DIS to include
the appropriate threshold behaviour in coefficient functions at each 
order \cite{ACOT}. J.~Bartels also gave a summary of the contribution of higher 
twist operators
at small $x$ \cite{bartels}. Within the context of the double-leading-log 
approximation, and having to model the inputs, the contributions of the 
four sources due to gluon operators show a potentially large
cancellation in $F_2$, but imply a large negative twist four correction to
$F_L$.  

S.~Moch gave a summary of NNLO
calculations of splitting functions and coefficient functions \cite{mocha}.
These rely on calculating the Mellin moments of the structure functions,
which results in simplification since internal propagators in diagrams 
which depend on the parton momentum $p$ can be expanded in powers 
of $(p\cdot q/q^2)^N$ where $N$ indeed corresponds to the Mellin moment 
variable. For the diagrams with only one internal line dependent on $p$,
known as basic building blocks, this reduces 4-point diagrams to 2-point 
diagrams and the
calculation is greatly simplified. Various techniques then also have to be
used to relate more complicated diagrams to these building blocks
\cite{mochb}.     
A number of fixed moments of 3-loop splitting functions and coefficient 
functions have already been calculated \cite{NNLOmoms}. The complete 
calculation of non-singlet quantities is nearly finished.
The much more complicated singlet quantities will be a little longer. 

Finally, we had an update on both the MRST  and CTEQ 
parton distributions~\cite{mrst01,cteq6}.
W.K.Tung~\cite{tung} concentrated on the treatment of uncertainties
due to experimental errors (discussed in Sec.~\ref{sec:qcderr}).  
R.~Thorne~\cite{thorne} instead emphasized the need to
understand theory errors for partons as well as the development of
experimental errors. For example, MRST have used the approximate NNLO 
splitting functions in \cite{NNLOsplit} to perform global fits and
make predictions at NNLO \cite{MRSTNNLO}.
This suggests that NNLO leads to a bigger correction even to the 
W cross-section at
Tevatron than the experimental error within NLO. The theory errors 
associated with higher orders are probably much bigger for gluon
dominated quantities. Additionally, detailed investigation of cuts
on data \cite{cuts} suggest that the fits improve if the lowest $Q^2$ 
and particularly lowest $x$ data are cut out, 
and the predictions for cross-sections with the 
new partons change. This suggests 
potentially large corrections to NLO DGLAP at low $Q^2$ and low $x$.

\subsection{Theoretical Conclusions}

As outlined, there are various different approaches to calculating 
structure functions, and there 
has been real progress in some of these calculations, e.g. NNLO in the usual 
expansion in $\alpha_S$, in the colour glass model and in NLO corrections to
BFKL impact factors. All the approaches are probably applicable in their 
own regime, and in some cases can be extrapolated with considerable success
for surprising distances. However, this success may sometimes lead to 
unwarranted claims that one approach is actually particularly appropriate. 
There needs to be more real understanding of precisely
where the regimes are and how they can be combined in order to produce the 
best overall theory with the maximum predictive power. In our opinion the 
best theory, 
particularly when one considers the predictive power for a 
wide range of processes over a range of different experiments (HERA, 
Tevatron, NuTev, LEP, LHC) is probably the collinear factorization 
theorem, but 
improved as much as possible by, for example, resummations at small and 
large $x$, higher twist corrections, etc. Clearly this will then require 
modification in the nonperturbative regime. This construction of the best 
complete, universally applicable theory is a difficult task, and help will
be needed from even more precise and wide-ranging data.    

\section{Conclusions}

New precise results on structure functions across a wide range of
$Q^2$, from $0.35$ to $30000 \GeV^2$, have been presented this year. 
The data are now systematics rather than
statistics limited such that QCD fits to extract parton distributions and
$\alpha_s$ have to consider correlated systematic errors. The precision of
the data requires extension of the conventional formalism of NLO QCD, as 
embodied in the DGLAP formalism, and this challenge 
is being met as this formalism extended in various directions: to NNLO, to
small $x$, to high density and to the non-perturbative regime.

\end{document}